\documentclass[iop]{emulateapj}
\usepackage{natbib}
\usepackage{color}
\usepackage[encapsulated]{CJK}
\usepackage{ucs}
\usepackage[utf8x]{inputenc}
\usepackage{gensymb}
\usepackage{ulem}
\usepackage{amsmath}
\usepackage{amssymb}

\newenvironment{Contfigure*}{%
\renewcommand{\thefigure}{\arabic{figure} (Cont.)}%
\addtocounter{figure}{-1}%
\begin{figure*}}{%
\renewcommand{\thefigure}{\arabic{figure}}%
\end{figure*}}

\begin{document}
\begin{CJK}{UTF8}{gbsn}

\title{\textrm{C}~\textsc{iii}] Emission in Star-forming Galaxies at $z\sim1$}

\author{Xinnan Du (杜辛楠)\altaffilmark{1}, Alice E. Shapley\altaffilmark{1}, Crystal L. Martin\altaffilmark{2}, Alison L. Coil\altaffilmark{3}}

\altaffiltext{1}{Department of Physics and Astronomy, University of California, Los Angeles CA, 90095, USA}

\altaffiltext{2}{Department of Physics, University of California, Santa Barbara CA, 93106, USA}

\altaffiltext{3}{Department of Physics, University of California, Santa Diego CA, 92093, USA}

\slugcomment{Draft Version \today}

\shorttitle{\textrm{C}~\textsc{iii} Emission}
\shortauthors{Du}

\begin{abstract}
The \textrm{C} \textsc{iii}]$\lambda\lambda$1907, 1909 rest-frame UV emission doublet has recently been detected in galaxies during the epoch of reionization ($z > 6$), with high equivalent width ($>10 \mbox{\AA}$, rest frame). Currently, it is possible to obtain much more detailed information for star-forming galaxies at significantly lower redshift. Accordingly, studies of their far-UV spectra are useful for understanding the factors modulating the strength of \textrm{C} \textsc{iii}] emission. We present the first statistical sample of \textrm{C} \textsc{iii}] emission measurements in star-forming galaxies at $z\sim1$. Our sample is drawn from the DEEP2 survey and spans the redshift 0.64 $\leqslant z \leqslant$ 1.35 ($\langle z \rangle$ =1.08). We find that the median equivalent width (EW) of individual \textrm{C} \textsc{iii}] detections in our sample (1.30 $\mbox{\AA}$) is much smaller than the typical value observed thus far at $z>6$. Furthermore, out of 184 galaxies with coverage of \textrm{C} \textsc{iii}], only 40 have significant detections. Galaxies with individual \textrm{C} \textsc{iii}] detections have bluer colors and lower luminosities on average than those without, implying that strong \textrm{C} \textsc{iii}] emitters are in general young and low-mass galaxies without significant dust extinction. Using stacked spectra, we further investigate how \textrm{C} \textsc{iii}] strength correlates with multiple galaxy properties ($\mbox{M}_{B}$, $\ub$, $\mbox{M}_{*}$, star-formation rate, specific star-formation rate) and rest-frame near-UV (\textrm{Fe} \textsc{ii}* and \textrm{Mg} \textsc{ii}) and optical ([\textrm{O} \textsc{iii}] and $\mbox{H}\beta$) emission line strengths. These results provide a detailed picture of the physical environment in star-forming galaxies at $z\sim1$, and motivate future observations of strong \textrm{C} \textsc{iii}] emitters at similar redshifts.

\end{abstract}


\section{INTRODUCTION}
\label{sec:Intro}

Rest-frame UV spectroscopy contains rich information regarding the physical properties of the interstellar medium (ISM) in star-forming galaxies. The abundant nebular emission features in the far-UV, including \textrm{C}~\textsc{iii}]$\lambda\lambda$1907, 1909, \textrm{C}~\textsc{iv}$\lambda\lambda$1548, 1550, \textrm{He}~\textsc{ii}$\lambda$1640, \textrm{O}~\textsc{iii}]$\lambda\lambda$1661, 1666 and \textrm{Si}~\textsc{iii}]$\lambda\lambda$1882, 1892, are especially important in probing the ionized gas in \textrm{H}~\textsc{ii} regions. These lines can be combined to infer various physical parameters, such as metallicity, abundance pattern, and ionization parameter, and provide valuable constraints on the ionizing stellar populations.

One of the strongest far-UV emission lines is typically the \textrm{C}~\textsc{iii}]$\lambda\lambda$1907, 1909 doublet. Given its prominence and detectability in bright, star-forming galaxies at $z\sim6-7$, \textrm{C}~\textsc{iii}] has been suggested as an alternative to Ly$\alpha$ emission for estimating redshifts during the reionization epoch \citep{Stark2014,Stark2015}. Indeed, spectroscopic confirmation is crucial yet challenging for $z>6$ galaxy candidates, in which Ly$\alpha$ is significantly attenuated by the neutral intergalactic medium (IGM) \citep[e.g.,][]{Vanzella2011,Treu2012,Treu2013,Schenker2014,Pentericci2014,Tilvi2014,Caruana2014}. In contrast, nebular \textrm{C}~\textsc{iii}] photons are negligibly affected by IGM absorption.

Nebular \textrm{C}~\textsc{iii}] emission has been studied both in the local universe \citep[][Pena-Guerrero et al. 2015]{Gia1996,Leitherer2011,Berg2016} and at redshift $>2$ \citep{Shapley2003,Erb2010,Bayliss2014,Stark2014,Stark2015,Rigby2015,Zitrin2015,Stark2016,Debarros2016,Vanzella2016,Steidel2016}. Large rest-frame EWs ($>5\mbox{\AA}$) of \textrm{C}~\textsc{iii}] have been detected at all these redshifts, and appear common at $z\gtrsim6-7$ \citep[$\gtrsim10\mbox{\AA}$;][]{Stark2015,Stark2016}. It has been suggested that galaxies showing strong \textrm{C}~\textsc{iii}] emission are in general blue, faint and low-mass, with high specific star-formation rates \citep[sSFRs;][]{Stark2014}. Furthermore, results from photoionization models indicate that \textrm{C}~\textsc{iii}] emission is enhanced at lower gas-phase metallicity, higher ionization parameters and harder radiation fields \citep[e.g.,][]{Erb2010,Stark2015,Stark2016,Berg2016,Jaskot2016}. Many studies have also shown that the strength of \textrm{C}~\textsc{iii}] correlates with that of other emission features, such as Ly$\alpha$ \citep{Shapley2003,Stark2014, Stark2015,Rigby2015,Jaskot2016} and [\textrm{O}~\textsc{iii}]$\lambda$5007 \citep{Jaskot2016}, in the sense that large \textrm{C}~\textsc{iii}] EW tends to appear in galaxies with strong Ly$\alpha$ and [\textrm{O}~\textsc{iii}] emission.

In principle, strong \textrm{C}~\textsc{iii}] emitters at $z>6$ are ideal sources to study how \textrm{C}~\textsc{iii}] emission relates in general to the properties of host galaxies and other emission lines. However, due to the poor data quality and small current sample of these extreme \textrm{C}~\textsc{iii}] emitters, it is exceptionally challenging to conduct studies in a statistical manner. Another way of dissecting the $z>6$ \textrm{C}~\textsc{iii}] emitters is by studying lower-redshift analogs. It is more straightforward at lower redshift to assemble multi-wavelength data over a wide rest-frame wavelength range and collect rest-UV spectra with higher signal-to-noise. Therefore, targeting lower-redshift systems enables us to study in detail the factors that modulate the strength of \textrm{C}~\textsc{iii}]. 

Here we analyze the nebular \textrm{C}~\textsc{iii}] emission transition for the first time at $z\sim1$. One of our goals is to investigate whether the correlations between \textrm{C}~\textsc{iii}] and a variety of galaxy properties, seen both at lower and higher redshifts, are redshift-independent (i.e., that they also remain true at $z\sim1$). Our $z\sim1$ sample can potentially connect the local and $z>2$ universe in redshift space and provide a complete understanding of the factors affecting the production of nebular \textrm{C}~\textsc{iii}] emission. More importantly, we aim to infer the physical conditions in typical star-forming galaxies at $z\sim1$ from the characteristic strength of \textrm{C}~\textsc{iii}], based on both observational constraints and photoionization models.

We provide a brief overview of the observations and data reduction, and discuss the properties of the \textrm{C}~\textsc{iii}] sample in Section~\ref{sec:data}. In Section~\ref{sec:meas}, we describe the measurement of the \textrm{C}~\textsc{iii}] emission profile in our data. We present the \textrm{C}~\textsc{iii}] EW distribution in Section~\ref{sec:res}, along with the correlations of the \textrm{C}~\textsc{iii}] strength with multiple galaxy properties (i.e., $\mbox{M}_{B}$, $\ub$, $\mbox{M}_{*}$, SFR, sSFR), the strength of the near-UV \textrm{Fe}~\textsc{ii}* and \textrm{Mg}~\textsc{ii} emission features, and gas-phase metallicity. Finally, we discuss and summarize our results in Sections \ref{sec:dis} and \ref{sec:sum}, respectively.

\begin{figure*}
\includegraphics[width=1.0\linewidth]{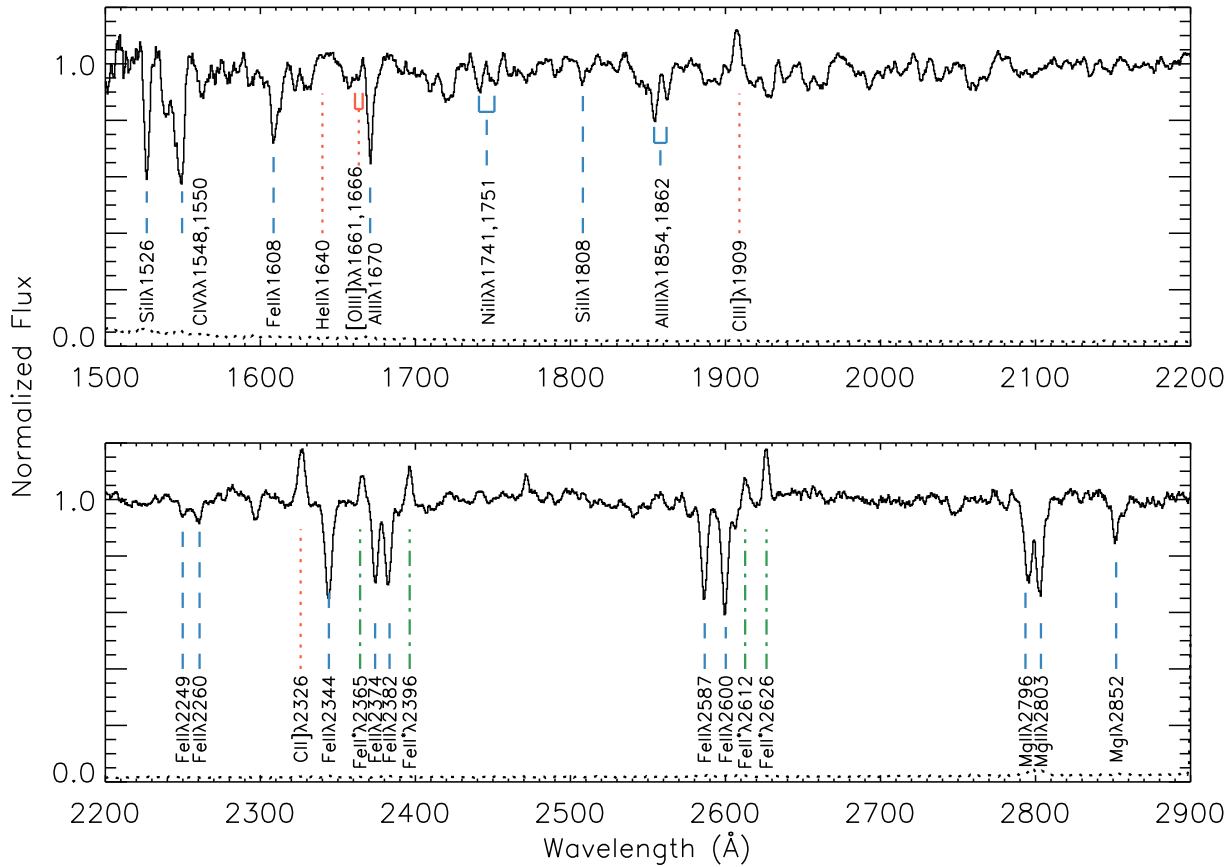}
\caption{Composite continuum-normalized UV spectrum (black solid line) constructed from 184 objects with \textrm{C}~\textsc{iii}] coverage in the DEEP2/LRIS sample. The composite spectrum was created using median stacking, with the noise level (black dotted line) estimated from Monte Carlo methods, as described in Section \ref{sec:galprop}. Characteristic emission and absorption features are identified, with the following labels: interstellar absorption transitions (blue dashed lines), fine structure transitions (green dash-dot lines) and nebular emission transitions (red dotted lines). At the resolution of the LRIS spectra included in this composite (i.e., 400-line spectra, and 600-line spectra smoothed to match the 400-line resolution), the \textrm{C}~\textsc{iii}] doublet is blended.}
\end{figure*}
\label{fig:spec}

Throughout this paper, we adopt a standard $\Lambda$CDM model with $\Omega_{m}=0.3$, $\Omega_{\Lambda}=0.7$ and $H_{0}=$70 km $\mbox{s}^{-1}$. All wavelengths are measured in vacuum. Magnitudes and colors are on the AB system.

\begin{figure*}
\includegraphics[width=0.5\linewidth]{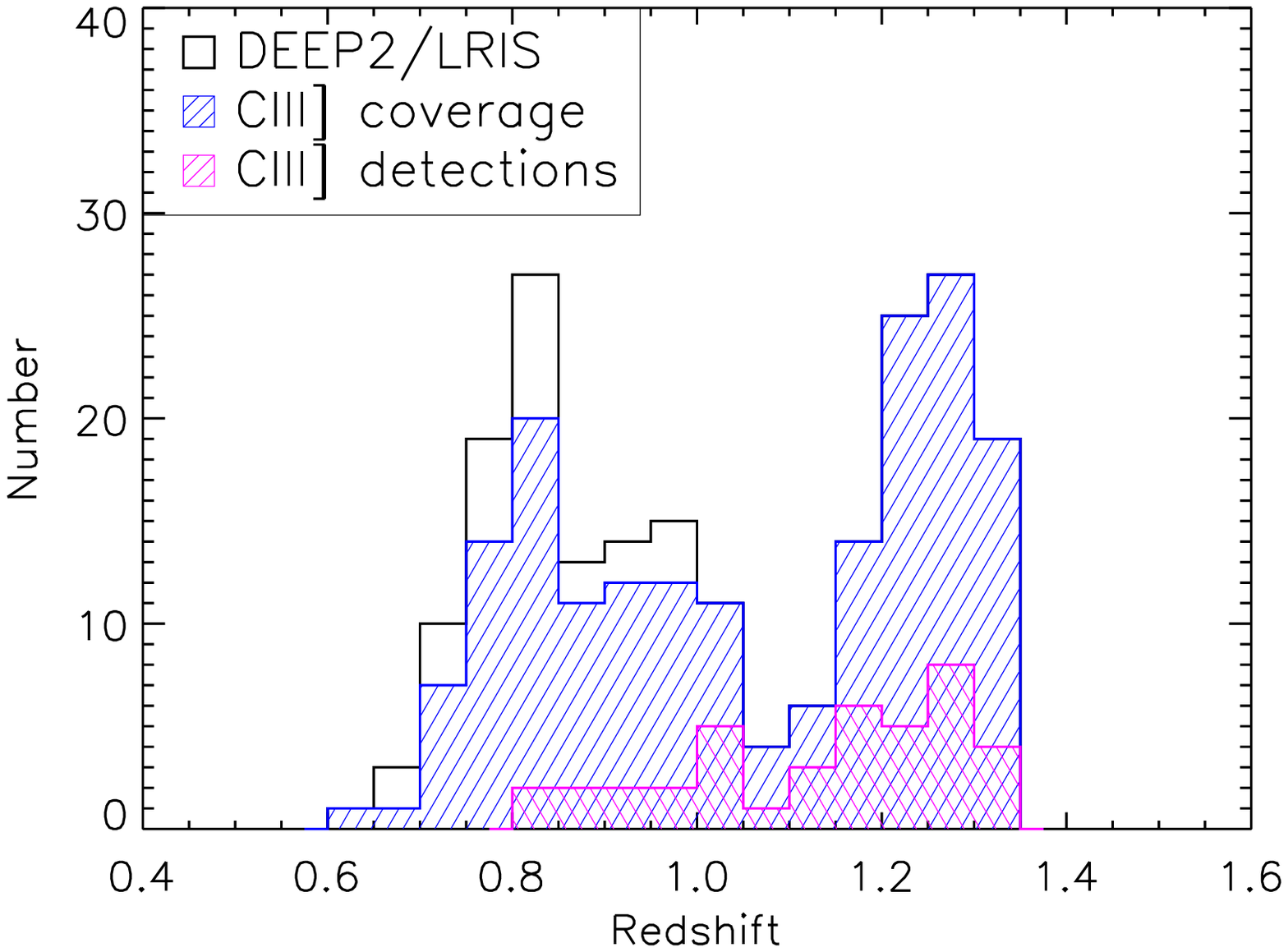}
\includegraphics[width=0.5\linewidth]{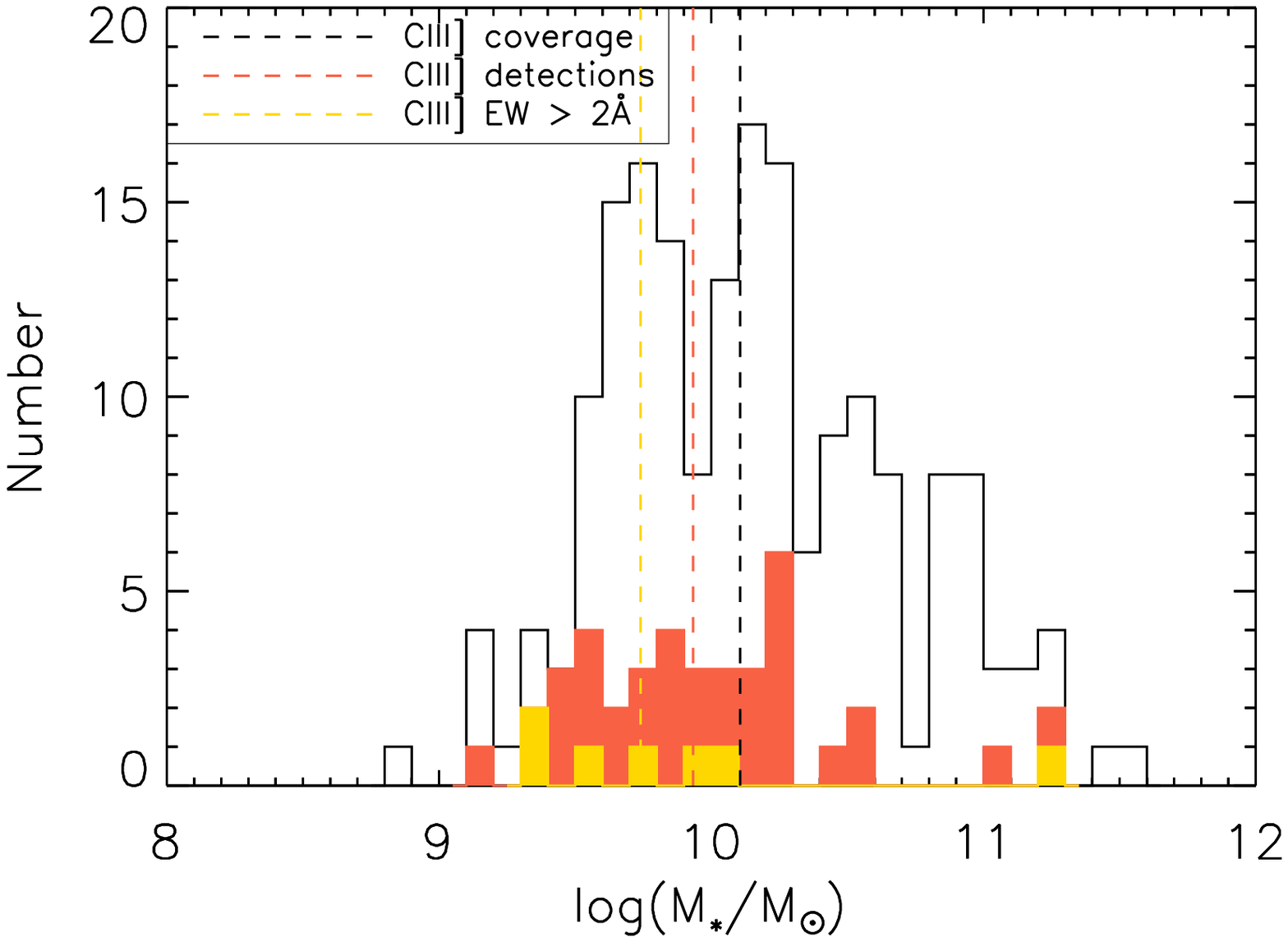}
\begin{center}
\includegraphics[width=0.8\linewidth]{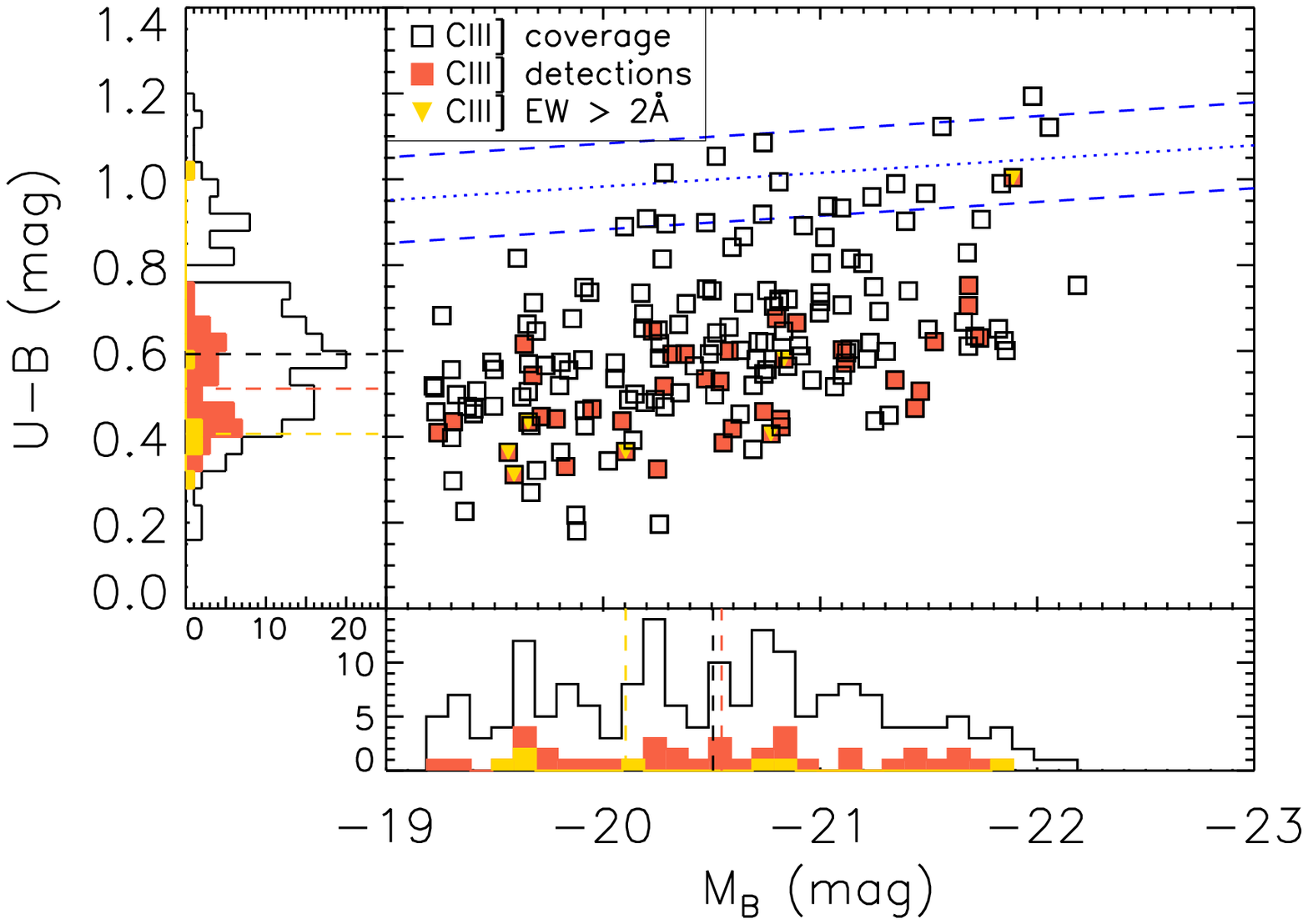}
\caption{Properties of the galaxies in the DEEP2 \textrm{C}~\textsc{iii}] sample. \textbf{Top left:} Redshift distribution. The black open bar represents the parent sample of 208 DEEP2/LRIS galaxies; the dashed blue bar represents 184 star-forming galaxies with \textrm{C}~\textsc{iii}] coverage; 
the dashed magenta bar represents 40 galaxies meeting the criterion of $>$ 3$\sigma$ \textrm{C}~\textsc{iii}] EW detection in the rest frame. \textbf{Top right:} Stellar mass distribution. The black histogram shows the full sample with \textrm{C}~\textsc{iii}] coverage (184 objects); the red and yellow filled histograms show, respectively, objects with significant \textrm{C}~\textsc{iii}] detections (40 objects) and rest-frame EW $>2\mbox{\AA}$ (7 objects). The median values of these different samples are shown with dashed vertical lines in corresponding colors. \textbf{Bottom:} $\ub$ vs. $\mbox{M}_{B}$ color-magnitude diagram with histograms in $\ub$ color and $B-$ band absolute magnitude. Squares represent 184 objects with \textrm{C}~\textsc{iv} coverage; red filled squares represent those with \textrm{C}~\textsc{iii}] $S/N>3$ (40 objects); yellow triangles represent 7 objects with \textrm{C}~\textsc{iii}] rest-frame EW measurements $> 2 \mbox{\AA}$.
The dotted line marks the division between the ``red sequence" and the ``blue cloud" at $z\sim1$ in the DEEP2 sample \citep{Willmer2006}. Legends for $\ub$ and $\mbox{M}_{B}$ histograms are the same as in the top right panel.}
\label{fig:galprop}
\end{center}
\end{figure*}

\section{Observations, Data Reduction and Sample}
\label{sec:data}

A full description of our $z\sim1$ dataset has been presented in \citet{Martin2012}. We summarize the relevant details in this section. Objects presented in this paper were drawn from the Deep Extragalatic Evolutionary Probe 2 \citep[DEEP2;][]{Newman2013} galaxy redshift survey and observed with the Low Resolution Imager and Spectrometer \citep[LRIS,][]{Oke1995,Steidel2004} on the Keck I telescope. The LRIS data were collected during 4 observing runs from 2007 to 2009 using 9 multi-object slitmasks with $1.2''$ slits. A total of 208 $B<24.5$ galaxies were observed in the redshift range $z=0.4-1.4$ with $\langle z \rangle=1.01$, based on apparent magnitude $B<24.5$. We used two configurations for the LRIS observations: the primary set-up (6 masks) featured the 400 lines $\mbox{mm}^{-1}$ grism on the blue side with an average effective resolution of 435 $\mbox{km s}^{-1}$ full width at half-maximum (FWHM), the d680 dichroic, and the 831 lines $\mbox{mm}^{-1}$ grating on the red side with FWHM of 150 $\mbox{km s}^{-1}$ ($R=700$). The secondary configuration (3 masks), which was mainly designed to obtain near-UV spectroscopy of brighter galaxies under less optimal observing conditions, had the 600 lines $\mbox{mm}^{-1}$ grism on the blue side with FWHM of 282 $\mbox{km s}^{-1}$, the d560 dichroic, and the 600 lines $\mbox{mm}^{-1}$ red grating with FWHM of 220 $\mbox{km s}^{-1}$ ($R=1100$). The primary set-up enabled the coverage of the rest-UV spectra of our targets on the blue side, and the [\textrm{O}~\textsc{ii}] nebular emission doublet (a systemic redshift diagnostic) on the red side. For 600-line masks, we aimed for continuous wavelength coverage between blue and red side spectra, across the dichroic. The integration time ranged from 5 to 9 hours for most 400-line masks and was typically shorter (3-5 hours) for the 600-line masks.

All two-dimensional spectra were flat fielded, cleaned of cosmic rays, background-subtracted, extracted into one dimension (1D), wavelength and flux calibrated, and transformed to the rest frame. We utilized rest-frame $B$-band luminosities and $\ub$ colors measured from \citet{Willmer2006}, and stellar masses derived SED fitting with $BRIK$ photometry, assuming a \citet{Chabrier2003} IMF and \citet{Bruzual2003} spectral templates (see \citet{Bundy2006} for a full description). We also adopted $\mbox{SFR}_{UV}$, the dust-corrected SFR derived from $Galaxy$ $Evolution$ $Explorer$ (GALEX) UV measurements in our work. The spectral slope $\beta$, which characterizes the continuum flux ($f_{\lambda} \propto \lambda^{\beta}$) over the rest-frame wavelength range of $1250 - 2500\mbox{\AA}$, was used to determine the dust extinction in the UV based on the relationship from from \citet{Seibert2005}. The $\mbox{SFR}_{UV}$ values presented in this paper have been converted to a Chabrier IMF. 42 out of 184 objects in our sample fall in the AEGIS field, where the $\mbox{SFR}_{UV}$ measurements are available. For a detailed discussion of the DEEP2/LRIS galaxy properties, we refer the reader to \citet{Martin2012} and \citet{Kornei2012}.

The DEEP2/LRIS data are unique in that the masks were designed to have the coverage of the far-UV spectral region at $z\sim1$. As shown in the composite spectra (Figure~\ref{fig:spec}), multiple far-UV nebular emission features were covered in the LRIS data. These include \textrm{C}~\textsc{iv}$\lambda\lambda$1548, 1550, \textrm{He}~\textsc{ii}$\lambda$1640, \textrm{O}~\textsc{iii}]$\lambda\lambda$1661, 1666 and \textrm{C}~\textsc{iii}]$\lambda\lambda$1907, 1909, which can be potentially modeled to provide a rich window into the physical properties of the galaxies in our sample. Among these far-UV features, \textrm{C}~\textsc{iii}] is the strongest emission transition, and the one detected in the largest number of individual spectra. Therefore, we focus the bulk of our analysis on nebular \textrm{C}~\textsc{iii}] emission. We identified 186 out of 208 galaxies with coverage of \textrm{C}~\textsc{iii}]. Of these, we removed 2 objects (12015320 and 22028607) with obvious spectroscopic evidence of activity from an active galactic nucleus (AGN), in the form of [\textrm{Ne}~\textsc{v}]$\lambda3425$ emission. The remaining 184 objects comprise the first statistical sample of \textrm{C}~\textsc{iii}] at $z\sim1$. Among these 184 objects, 144 were observed on 400-line masks, and 40 on 600-line masks. Figure~\ref{fig:spec} shows a composite rest-UV spectrum created by stacking the continuum-normalized spectra of the sample of 184 galaxies with \textrm{C}~\textsc{iii}] coverage. Spectra were normalized as described in \citet{Du2016}. The strongest interstellar emission and absorption features are labeled.

We plot the galaxy properties (i.e., redshift, stellar mass, $\ub$ color and $B$-band absolute magnitude) of the sample of 184 galaxies with \textrm{C}~\textsc{iii}] coverage in Figure \ref{fig:galprop}. These objects range in redshift from 0.64 to 1.35 with a median of 1.08, $B$-band absolute magnitude ($\mbox{M}_{B}$) from $-18.31$ to $-22.19$ with a median of $-20.51$, stellar mass ($\log(\mbox{M}_{*}/\mbox{M}_{\sun}$)) from 8.88 to 11.58 with a median of 10.10, $\ub$ color from 0.18 to 1.19 with a median of 0.59 and $\mbox{SFR}_{UV}$ from 2 to 97 $\mbox{M}_{\sun}$ $\mbox{yr}^{-1}$ with a median of 13 $\mbox{M}_{\sun}$ $\mbox{yr}^{-1}$. Galaxies in our sample are predominantly star-forming galaxies in the ``blue cloud" \citep{Faber2007} at $z\sim1$.

\section{Measurements}
\label{sec:meas}

In this section, we describe the measurement of the strong far-UV nebular emission doublet, \textrm{C}~\textsc{iii}]$\lambda\lambda1907, 1909$. The \textrm{C}~\textsc{iii}] doublet is a collisionally excited forbidden/semi-forbidden transition, which is typically observed in \textrm{H}~\textsc{ii} regions in star-forming galaxies. The \textrm{C}~\textsc{iii}] doublet ratio is determined by the electron density. In the low-density regime, which we believe is the case for galaxies in our sample, [\textrm{C}~\textsc{iii}]$\lambda1907$ is stronger than \textrm{C}~\textsc{iii}]$\lambda1909$, and $I_{[\textrm{C}~\textsc{iii}]\lambda1907}/I_{\textrm{C}~\textsc{iii}]\lambda1909}$ asymptotes to a value of 1.55 in the low-density limit \citep{Oster2006}.

We continuum normalized the rest-frame UV calibrated 1D LRIS spectra using multiple discrete wavelength ranges (`windows') that are clean of spectral features, as defined by \citet{Rix2004}. Based on these windows, we modeled the continuum for all 184 spectra with \textrm{C}~\textsc{iii}] coverage using the IRAF $continuum$ routine, with a $spline3$ function of order $=8$. In cases where the fitted continuum level did not provide a good description of the observed spectrum due to the limited coverage of windows from \citet{Rix2004}, 
additional windows customized for each object were added to keep the fitted continuum 
reasonable and make sure that it scattered around unity. Monte Carlo simulations indicate
that the systematic uncertainty at each wavelength associated with continuum fitting is less than the corresponding 1$\sigma$ error spectrum, and therefore does not dominate the uncertainties in our EW measurements.

Given the different resolutions in the primary and secondary masks, the \textrm{C}~\textsc{iii}] doublet is blended in the 400-line spectra at the rest-frame wavelength while in principle resolved in the 600-line spectra. Accordingly, we fit the \textrm{C}~\textsc{iii}] emission with a single Gaussian profile for the 400-line spectra, and deblended the feature into two Gaussians for those observed with the 600-line masks. 

We used the IDL program MPFIT \citep{Mark2009} with the initial values 
of continuum flux level, line centroid, EW and Gaussian FWHM (for the overall blended profile in the 400-line spectra and/or for each deblended component in the 600-line spectra) estimated from the program $splot$ in IRAF. For the 600-line spectra, we fixed the doublet wavelengths at the rest-wavelength ratio, \textbf{$\lambda_{1907}/\lambda_{1909} = 1906.683/1908.734$,} and constrained the doublet EW ratio to vary between 1.0 to 1.55, as expected in the low-density regime.\footnote{Given that the doublet members are adjacent in wavelength, the doublet ratio in terms of intensity should be almost identical to that in terms of rest-frame EW.} The best fit was then determined where the $\chi^{2}$ of the fit reached a minimum.

We iterated the fitting over a narrower wavelength range for both 400-line and 600-line spectra: centroid$-4\sigma < \lambda < $centroid$+4\sigma$, where the centroid and $\sigma$ are, respectively, the returned central wavelength and standard deviation of the best-fit Gaussian profile from the initial MPFIT fit to the \textrm{C}~\textsc{iii}] doublet over $1898\mbox{\AA}$ to $1918\mbox{\AA}$. 
We then determined the significance of the \textrm{C}~\textsc{iii}] EW in each object and identified 40 out of 184 objects with a $>$ 3$\sigma$ \textrm{C}~\textsc{iii}] detection. Although we did not apply an explicit continuum $S/N$ threshold when determining the sample with significant \textrm{C}~\textsc{iii}] detections, all 40 galaxies have continuum $S/N>4$. We list the rest-frame \textrm{C}~\textsc{iii}] EW measurements for these 40 objects in Table \ref{tab:ciii_ew}, along with other basic galaxy properties.

\section{Results}
\label{sec:res}

In order to examine the factors that modulate the \textrm{C}~\textsc{iii}] emission profile and understand the physical nature of the gas probed by \textrm{C}~\textsc{iii}], we compare the distribution of \textrm{C}~\textsc{iii}] EWs in the DEEP2/LRIS sample with those in studies at other redshifts in Section \ref{sec:distr}. In addition, we investigate how \textrm{C}~\textsc{iii}] strength correlates with galaxy properties in both individual and composite spectra in Section \ref{sec:galprop}, and whether the strength of \textrm{C}~\textsc{iii}] displays any connections with those of other near-UV emission features in Section \ref{sec:nuv}. Finally, we explore the relation between \textrm{C}~\textsc{iii}] and metallicity in Section \ref{sec:metal}.

\subsection{\textrm{C}~\textsc{iii}] EW distribution}
\label{sec:distr}

\begin{figure}
\includegraphics[width=1.0\linewidth]{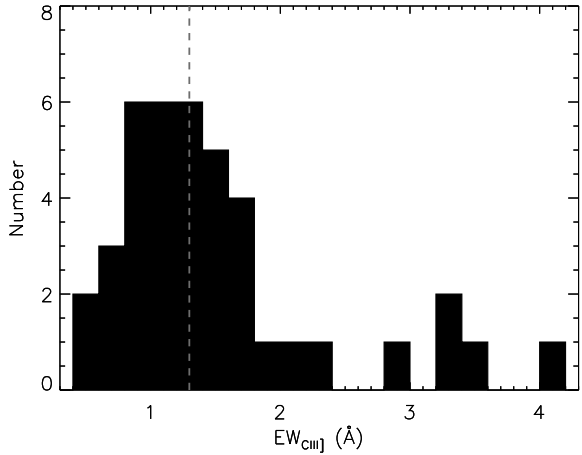}
\caption{\textrm{C}~\textsc{iii}] rest-frame EW distribution for the DEEP2/LRIS sample. Objects shown in the histogram have $\geqslant 3\sigma$ \textrm{C}~\textsc{iii}] EW detections. The gray dashed line marks the median EW, 1.30$\mbox{\AA}$.}
\label{fig:histr}
\end{figure}

\begin{deluxetable*}{ccccccccc}
\tablewidth{0pt}
  \tablecaption{Galaxy properties and $\textrm{C}~\textsc{iii}$] EW measurements}
  \tablehead{
    \colhead{ID} &
    \colhead{Redshift} &
    \colhead{R.A.} &
    \colhead{Dec.} &
    \colhead{$\mbox{M}_{B}$} &
    \colhead{$\ub$} &
    \colhead{$\log(\mbox{M}_{*})$} &
    \colhead{$\mbox{SFR}_{UV}$} &
    \colhead{$\mbox{EW}_{\textrm{C}~\textsc{iii}]}$} \\
    \colhead{} &
    \colhead{} &
    \colhead{(J2000)} &
    \colhead{(J2000)} &
    \colhead{(Magnitude)} &
    \colhead{(Magnitude)} &
    \colhead{($\mbox{M}_{\sun}$)} &
    \colhead{($\mbox{M}_{\sun} \mbox{yr}^{-1}$)} &
    \colhead{($\mbox{\AA}$)}  \\
    }
  \startdata
12008811 & 1.2156 & 14:16:55.33 & 52:30:24.91 & -20.80 &  0.68 & 10.16 &  17.98 & 1.3 $\pm$ 0.2 \\
12011619 & 1.0745 & 14:18:24.69 & 52:32:48.67 & -19.31 &  0.44 &  9.11 &   4.40 & 1.6 $\pm$ 0.5 \\
12012842 & 1.3148 & 14:16:56.46 & 52:33:13.77 & -21.68 &  0.75 & 11.09 & \nodata & 1.0 $\pm$ 0.3 \\
12013002 & 1.2184 & 14:16:50.04 & 52:33:46.59 & -20.23 &  0.65 &  9.56 &   8.51 & 1.1 $\pm$ 0.3 \\
12015682 & 1.2837 & 14:18:49.05 & 52:36:29.66 & -21.53 &  0.62 & 11.20 &  48.63 & 0.9 $\pm$ 0.3 \\
12015914 & 1.1046 & 14:18:22.11 & 52:35:27.05 & -19.83 &  0.33 & 10.11 &  15.85 & 0.9 $\pm$ 0.2 \\
12016050 & 0.9797 & 14:18:08.67 & 52:35:13.82 & -19.71 &  0.45 &  9.53 &  28.58 & 1.7 $\pm$ 0.5 \\
12016075 & 1.1174 & 14:18:05.99 & 52:34:08.43 & -19.68 &  0.54 &  9.41 &  10.78 & 1.5 $\pm$ 0.4 \\
12016903 & 1.1600 & 14:17:12.79 & 52:34:28.41 & -21.46 &  0.51 & 10.21 &  45.53 & 1.0 $\pm$ 0.2 \\
12019542 & 1.2785 & 14:18:49.95 & 52:40:22.16 & -21.68 &  0.71 & 10.40 &  72.25 & 0.8 $\pm$ 0.2 \\
22005715 & 1.2345 & 16:51:20.54 & 34:44:32.09 & -20.28 &  0.52 &  9.75 & \nodata & 1.3 $\pm$ 0.4 \\
22006207 & 1.2709 & 16:51:19.82 & 34:46:18.62 & -20.60 &  0.42 &  9.89 & \nodata & 1.1 $\pm$ 0.2 \\
22013182 & 0.8298 & 16:51:36.11 & 34:48:00.57 & -19.56 &  0.36 &  9.36 & \nodata & 2.3 $\pm$ 0.5 \\
22028986 & 1.1680 & 16:51:13.91 & 34:56:19.81 & -20.58 &  0.60 & 10.16 & \nodata & 1.0 $\pm$ 0.2 \\
22036194 & 1.1665 & 16:51:23.95 & 34:59:51.49 & -21.73 &  0.63 & 10.26 & \nodata & 0.5 $\pm$ 0.1 \\
22036688 & 1.1675 & 16:51:19.59 & 34:57:36.24 & -20.47 &  0.54 &  9.90 & \nodata & 1.7 $\pm$ 0.3 \\
22036975 & 0.9367 & 16:51:16.36 & 34:58:36.68 & -21.89 &  1.00 & 11.27 & \nodata & 4.1 $\pm$ 1.1 \\
22044809 & 1.1869 & 16:51:03.76 & 35:01:13.24 & -19.59 &  0.31 &  9.39 & \nodata & 3.6 $\pm$ 0.2 \\
22100920 & 1.2735 & 16:51:17.13 & 35:00:52.83 & -21.12 &  0.57 & 10.25 & \nodata & 0.7 $\pm$ 0.1 \\
22100930 & 1.2808 & 16:51:16.94 & 35:00:11.55 & -21.34 &  0.53 & 10.29 & \nodata & 1.3 $\pm$ 0.1 \\
32016857 & 0.9392 & 23:29:24.95 & 00:07:05.85 & -20.82 &  0.44 &  9.82 & \nodata & 1.7 $\pm$ 0.2 \\
32017018 & 1.1458 & 23:29:15.66 & 00:06:25.34 & -19.78 &  0.44 &  9.50 & \nodata & 1.0 $\pm$ 0.3 \\
32017112 & 1.0085 & 23:29:16.29 & 00:08:04.48 & -20.89 &  0.67 & 10.24 & \nodata & 0.9 $\pm$ 0.3 \\
32017188 & 1.2526 & 23:29:17.68 & 00:08:30.29 & -20.83 &  0.58 &  9.98 & \nodata & 3.2 $\pm$ 0.3 \\
32017272 & 1.0190 & 23:29:07.87 & 00:06:18.04 & -20.10 &  0.37 &  9.53 & \nodata & 2.8 $\pm$ 0.3 \\
32019861 & 1.3077 & 23:30:34.70 & 00:11:38.54 & -20.54 &  0.53 & 10.01 & \nodata & 1.5 $\pm$ 0.5 \\
32020062 & 1.2808 & 23:30:27.03 & 00:11:14.80 & -20.32 &  0.59 &  9.86 & \nodata & 1.3 $\pm$ 0.4 \\
32020384 & 1.2493 & 23:30:19.50 & 00:11:06.88 & -20.82 &  0.42 &  9.96 & \nodata & 1.1 $\pm$ 0.3 \\
32020728 & 1.0446 & 23:30:08.53 & 00:10:09.60 & -19.95 &  0.47 &  9.76 & \nodata & 1.8 $\pm$ 0.3 \\
32020769 & 1.3144 & 23:30:10.40 & 00:11:25.69 & -20.77 &  0.41 & 10.01 & \nodata & 2.2 $\pm$ 0.3 \\
42006875 & 0.8697 & 02:30:33.47 & 00:26:21.15 & -21.10 &  0.60 & 10.50 & \nodata & 1.4 $\pm$ 0.3 \\
42006904 & 1.0228 & 02:30:35.65 & 00:25:00.98 & -20.74 &  0.46 & 10.07 & \nodata & 0.6 $\pm$ 0.1 \\
42006915 & 0.8945 & 02:30:37.04 & 00:24:36.58 & -20.55 &  0.39 & 10.50 & \nodata & 1.0 $\pm$ 0.2 \\
42014618 & 1.0131 & 02:30:23.06 & 00:30:02.69 & -20.38 &  0.59 &  9.81 & \nodata & 0.5 $\pm$ 0.2 \\
42014653 & 0.8376 & 02:30:22.78 & 00:28:58.83 & -19.24 &  0.41 &  9.60 & \nodata & 1.6 $\pm$ 0.4 \\
42021266 & 0.9766 & 02:30:34.80 & 00:30:58.34 & -19.64 &  0.62 &  9.50 & \nodata & 1.3 $\pm$ 0.2 \\
42022173 & 1.3112 & 02:30:22.97 & 00:30:13.48 & -20.25 &  0.32 &  9.68 & \nodata & 1.5 $\pm$ 0.2 \\
42022307 & 1.2595 & 02:30:21.45 & 00:30:09.51 & -21.44 &  0.47 & 10.22 & \nodata & 0.8 $\pm$ 0.1 \\
42033338 & 1.2237 & 02:29:03.37 & 00:33:34.35 & -19.65 &  0.43 &  9.74 & \nodata & 3.4 $\pm$ 0.8 \\
42034223 & 1.2000 & 02:28:41.70 & 00:33:46.33 & -20.09 &  0.44 &  9.45 & \nodata & 1.3 $\pm$ 0.4 \\

 \enddata
\label{tab:ciii_ew}
\tablecomments{The $B$-band luminosity and $\ub$ color are taken from \citet{Willmer2006}. Stellar mass was estimated from SED fitting with $BRIK$ photometry, as presented in \citet{Bundy2006}. $\mbox{SFR}_{UV}$ is the dust-corrected UV SFR from GALEX measurements (full description in Section \ref{sec:data}). The typical uncertainty is $3\times10^{-4}$ in redshift, 0.1 - 0.2 dex in stellar mass, $\sim$0.05 mag in $\ub$ color, and $<0.1$ mag in $\mbox{M}_{B}$. Rest-frame \textrm{C}~\textsc{iii}] EWs were measured using Gaussian fits.}
\end{deluxetable*}

As described in Section \ref{sec:meas}, we found 40 objects with \textrm{C}~\textsc{iii}] EW $>$ $3\sigma$. Figure \ref{fig:histr} shows the rest-frame EW distribution of significant \textrm{C}~\textsc{iii}] detections in the DEEP2/LRIS sample. The \textrm{C}~\textsc{iii}] EW in this sample ranges from 0.5$\mbox{\AA}$ to 4.1$\mbox{\AA}$, with a median of 1.3$\mbox{\AA}$. In comparison with the \textrm{C}~\textsc{iii}] measurements in the low-mass, low-metallicity systems at high redshift \citep[$z>6$;][see Figure \ref{fig:z_ew}]{Erb2010, Stark2014, Stark2015, Zitrin2015, Stark2016, Vanzella2016, Debarros2016}, which have a median of $\gtrsim$ 10$\mbox{\AA}$, the EW of \textrm{C}~\textsc{iii}] at $z\sim1$ is noticeably smaller. We discuss the causes of the weak \textrm{C}~\textsc{iii}] emission and inferred physical environment in $z\sim1$ star-forming galaxies in Section \ref{sec:dis}.

\subsection{\textrm{C}~\textsc{iii}] vs. galaxy properties}
\label{sec:galprop}

\begin{figure*}
\includegraphics[width=0.5\linewidth]{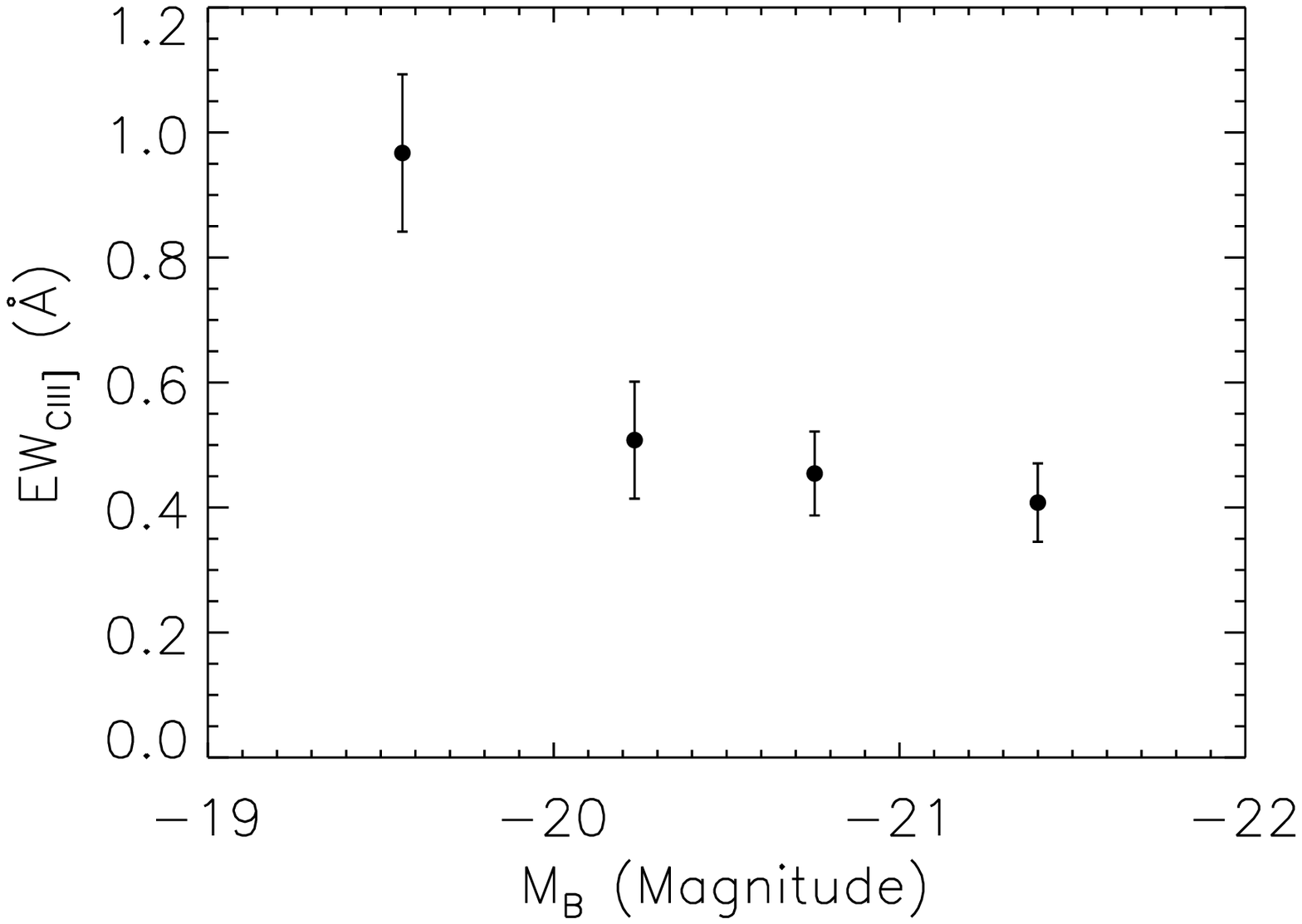}
\includegraphics[width=0.5\linewidth]{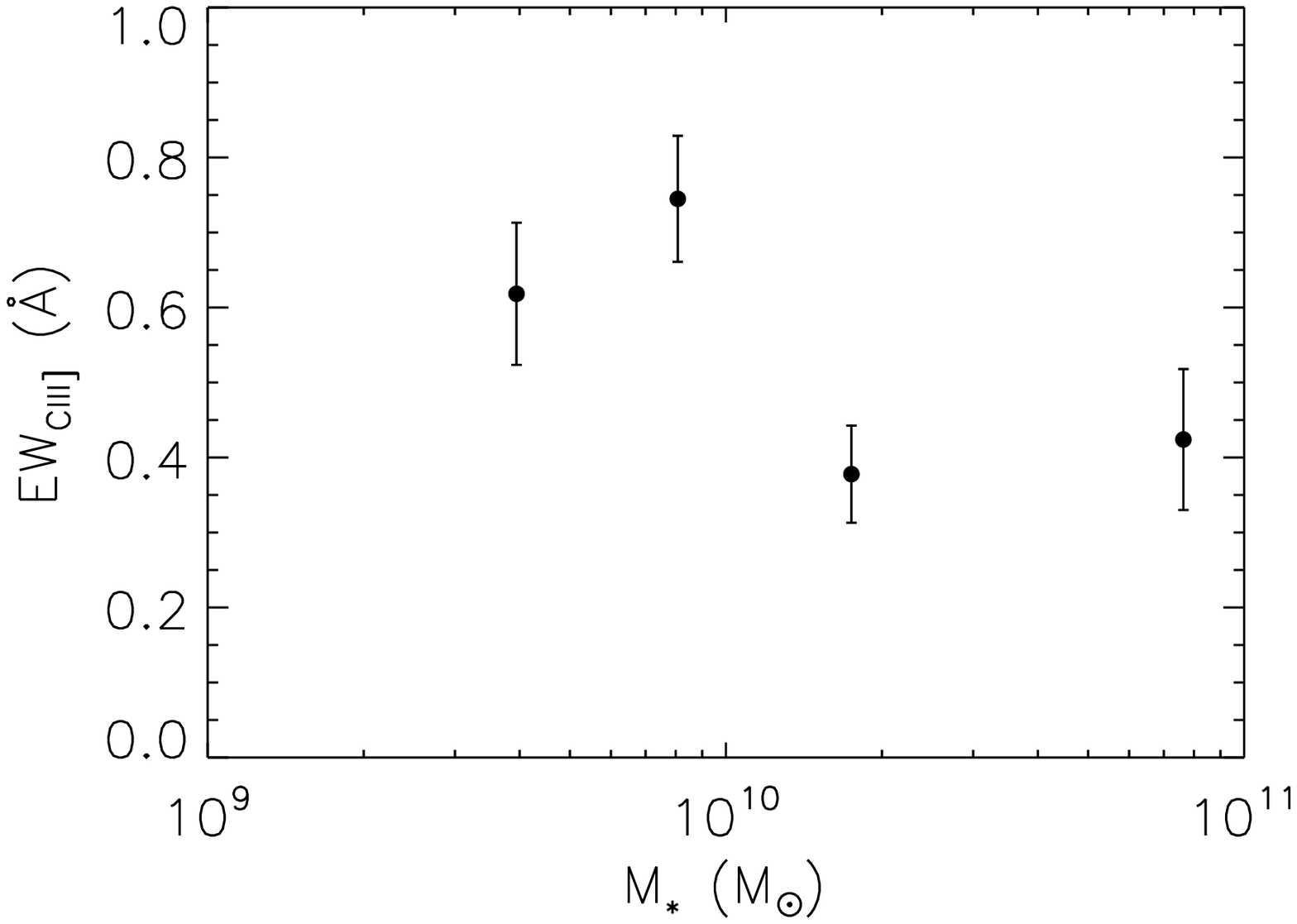}
\includegraphics[width=0.5\linewidth]{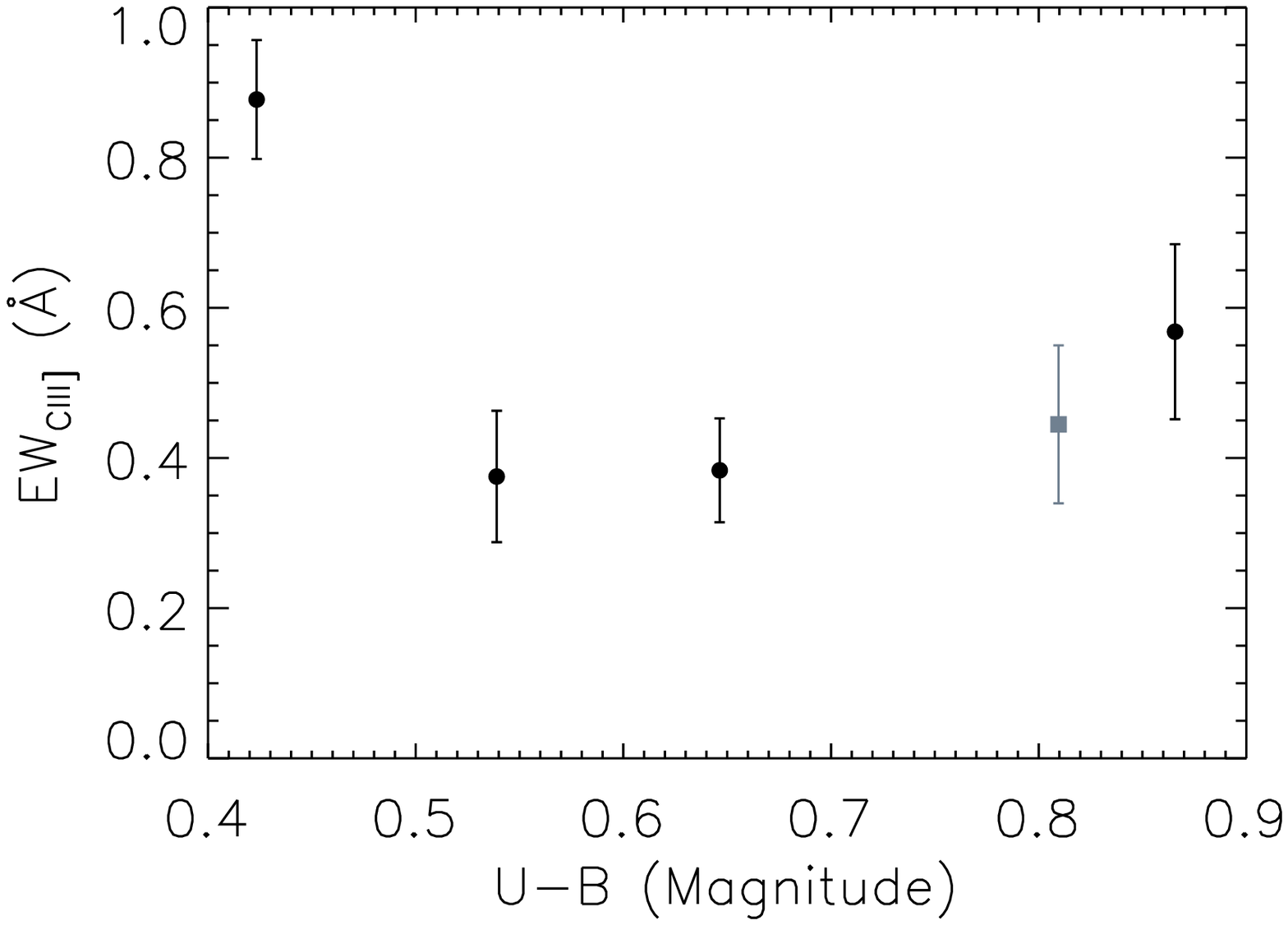}
\includegraphics[width=0.5\linewidth]{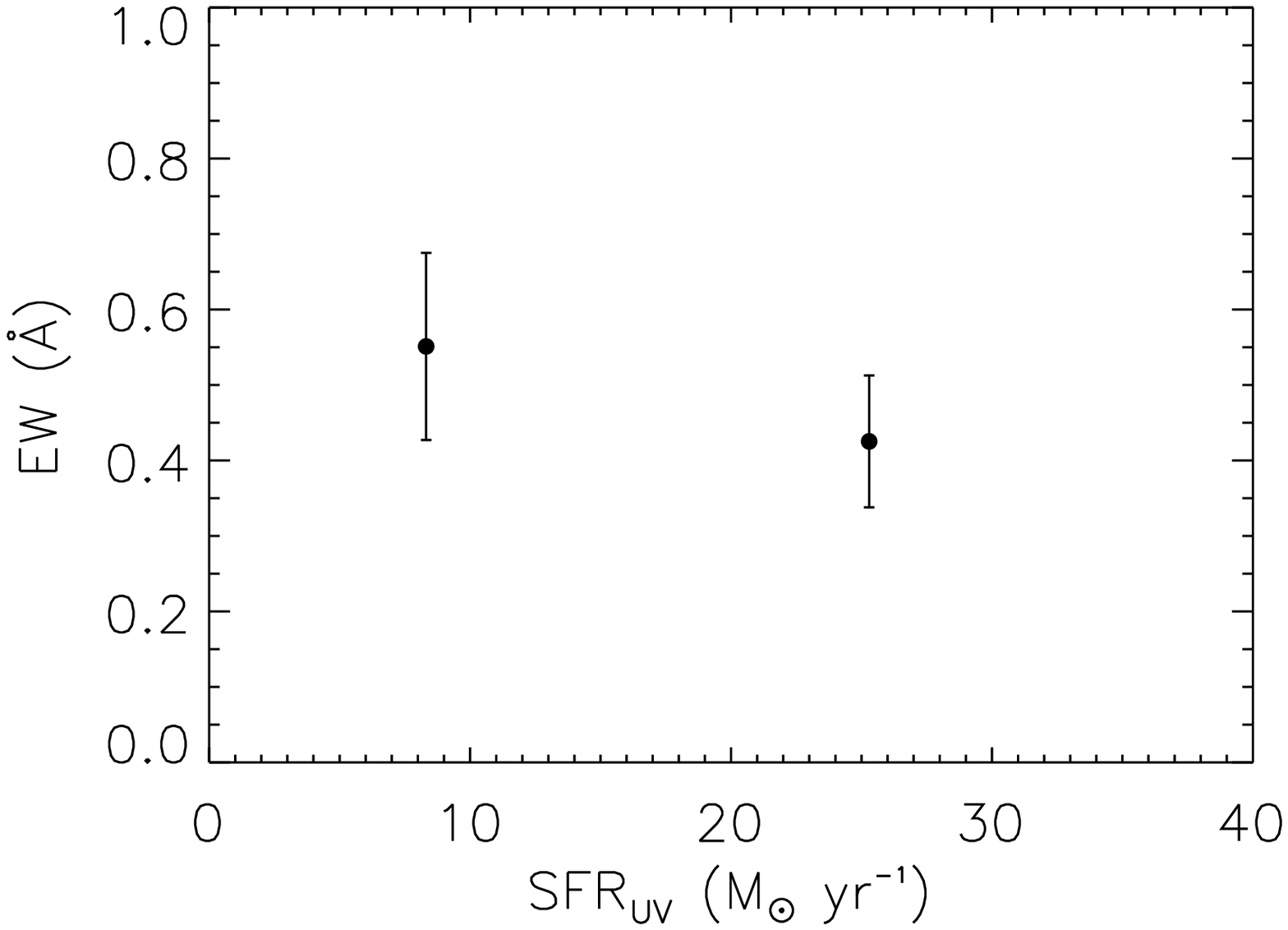}
\begin{center}
\includegraphics[width=0.5\linewidth]{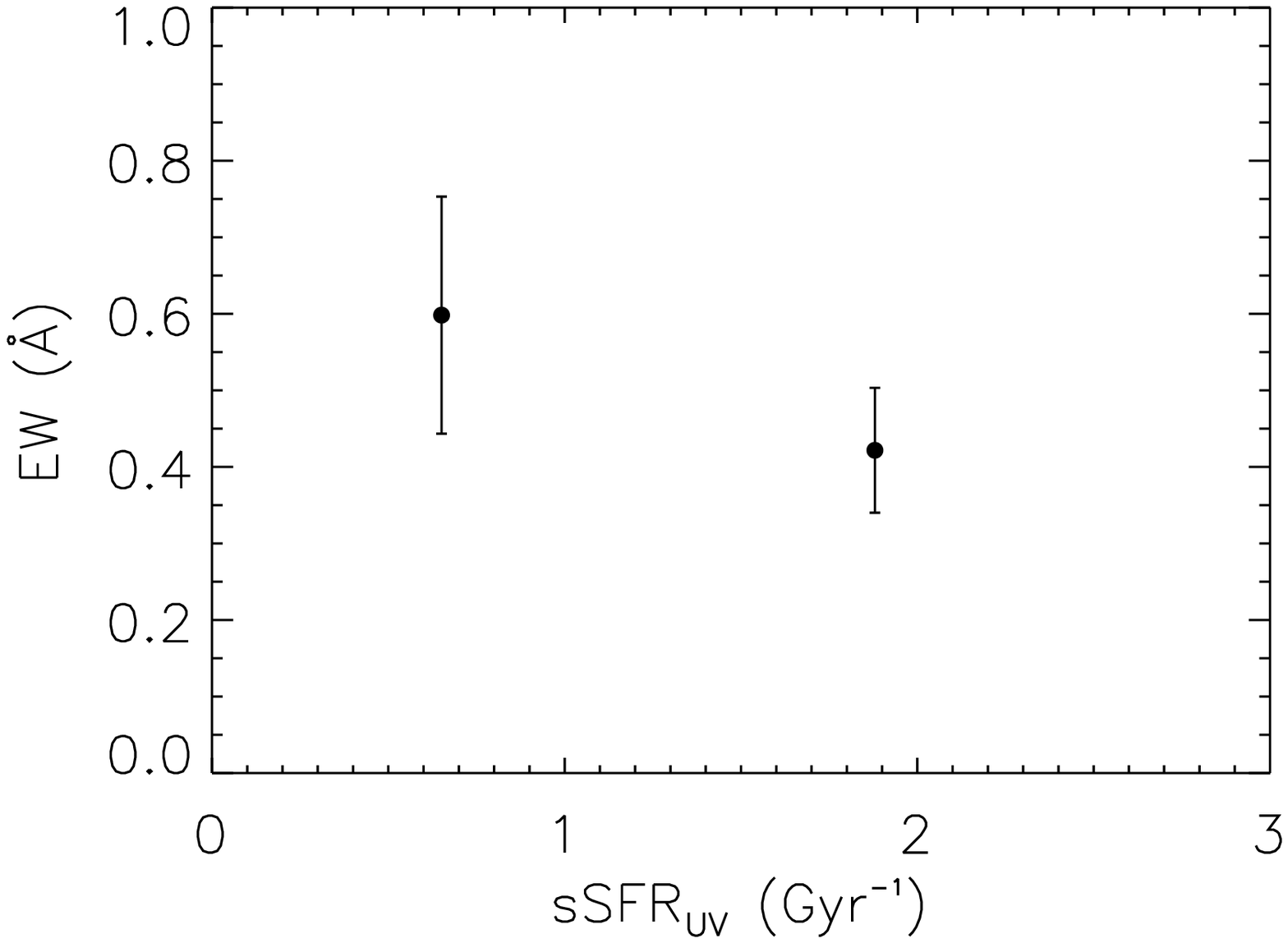}
\caption{\textbf{Top to bottom, left to right:} \textrm{C}~\textsc{iii}] EW vs. $B-$band absolute magnitude, stellar mass, $\ub$ color, $\mbox{SFR}_{UV}$, and $\mbox{sSFR}_{UV}$ in the composite spectra. The gray square in the middle left panel represents the \textrm{C}~\textsc{iii}] EW measurement from the reddest $\ub$ composite spectrum made out of a more controlled sample, where 32 objects with [\textrm{Ne}~\textsc{v}] coverage and insignificant ($\leqslant3\sigma$) [\textrm{Ne}~\textsc{v}] detections are included.}
\label{fig:stack_ew}
\end{center}
\end{figure*}

Studies of $z>2$ galaxies show that strong \textrm{C}~\textsc{iii}] emission tends to be observed in blue, faint, low-mass galaxies with large sSFRs \citep{Stark2014}. In order to examine if these correlations also apply to $z\sim1$, we investigate how galaxy properties relate to the \textrm{C}~\textsc{iii}] strength. For this analysis, we utilize the subsample of individual \textrm{C}~\textsc{iii}] detections and the composite spectra constructed from our full sample with \textrm{C}~\textsc{iii}] coverage (both individual detections and non-detections). We observe no apparent relations \textbf{($\lesssim 1.0\sigma$)} between \textrm{C}~\textsc{iii}] and either $\mbox{SFR}_{UV}$ or $\mbox{sSFR}_{UV}$ (middle right and bottom panels of Figure \ref{fig:stack_ew}). Therefore, we only highlight results regarding $B-$band absolute magnitude, $\ub$ color and stellar mass in this section.

We mark galaxies with individual \textrm{C}~\textsc{iii}] detections as red squares in the upper right and bottom panels of Figure \ref{fig:galprop}. These objects segregate from the full sample of \textrm{C}~\textsc{iii}] coverage in that they have bluer $\ub$ color and lower stellar mass. The median color and stellar mass of objects with significant \textrm{C}~\textsc{iii}] detections are $\ub = 0.51$ (0.08 magnitude bluer than that of the full sample) and $\log(\mbox{M}_{*}/\mbox{M}_{\sun}) = 9.93$ (0.18 dex smaller than that of the full sample), respectively. Moreover, objects with large \textrm{C}~\textsc{iii}] EW ($> 2\mbox{\AA}$) (7 objects, yellow triangles in Figure \ref{fig:galprop}), mainly occupy the lower-left corner in the color-magnitude diagram. These objects have a median $\ub$ color of 0.41 and a median stellar mass of $\log(\mbox{M}_{*}/\mbox{M}_{\sun}) = 9.74$, indicating that strong \textrm{C}~\textsc{iii}] emission is associated with blue and low-mass systems with little dust extinction. The only outlier in the color-magnitude diagram is the galaxy 22036975, which has the highest \textrm{C}~\textsc{iii}] EW ($4.09\mbox{\AA}$) in the sample of detections and stands out in having an unusual red color ($\ub=1.00$ mag) compared to other objects with \textrm{C}~\textsc{iii}] detections. In fact, this object falls in the ``green valley" between the blue cloud and red sequence (Figure \ref{fig:galprop}, bottom). One possible explanation of strong CIII] emission accompanied by a red color is the presence of an AGN. Indeed, both of the AGNs identified spectroscopically in our sample (see Section \ref{sec:data}) show green-valley colors and significant \textrm{C}~\textsc{iii}] emission. However, we find no evidence of an AGN in 22036975 based on the non-detection of the [\textrm{Ne}~\textsc{v}]$\lambda$3425 line. The 3$\sigma$ upper limit on the [\textrm{Ne}~\textsc{v}] rest-frame EW is 0.72 $\mbox{\AA}$ in 22036975, which is much lower than the typical EW observed in the optically-selected narrow-line AGNs \citep{Zakamska2003}. Therefore, although we are not certain about the physical explanation of the redder color shown by this object relative to other \textrm{C}~\textsc{iii}] detections, we still treat it as a star-forming galaxy and retain it in our sample.

While the connections between \textrm{C}~\textsc{iii}] emission and $\ub$ color and stellar mass are suggestive in the sample of individual detections, we cannot ignore the fact that these 40 objects only comprise $\sim20\%$ of the full sample with \textrm{C}~\textsc{iii}] coverage. In order to investigate the correlations between spectral and galaxy properties displayed by the entire sample with \textrm{C}~\textsc{iii}] coverage, we constructed composite spectra in bins of specific galaxy properties. Additionally, using composite spectra potentially enables the measurement of weak transitions that are difficult to detect on an individual basis due to the low continuum $S/N$ in a single spectrum.

We divided the set of 184 objects with \textrm{C}~\textsc{iii}] coverage into four bins in stellar mass, $\ub$ color and $B$-band absolute magnitude, and two bins in $\mbox{SFR}_{UV}$ and $\mbox{sSFR}_{UV}$, given that the $\mbox{SFR}_{UV}$ measurements are only available for 42 objects observed in the AEGIS field. Each bin in stellar mass, $\ub$, and $\mbox{M}_{B}$ contained nearly the same number of galaxies (i.e., $\sim46$). We combined individual continuum-normalized spectra to create composite spectra, smoothing the 600-line spectra to the resolution of the 400-line spectra. We then used the IRAF routine $scombine$ to extract the median value of the normalized spectra at each wavelength to create a normalized composite spectrum for each bin in galaxy properties. To create the corresponding composite error spectra that account for both sample variance and measurement uncertainty, we bootstrap-resampled each bin and perturbed each spectrum in the bootstrap sample according to its own error spectrum. The perturbed spectra in the bootstrap sample were then combined to create a new composite spectrum. The process was repeated 100 times and the standard deviation of these 100 fake composites at each wavelength was adopted as the value of the composite error spectrum for each bin.

We fit the \textrm{C}~\textsc{iii}] feature in the stacked spectra with a single Gaussian profile using MPFIT to measure the EW (i.e., the method described in Section \ref{sec:meas}). Figure \ref{fig:stack_ew} shows the behavior of \textrm{C}~\textsc{iii}] EW in the stacked spectra in terms of $B-$band absolute magnitude, stellar mass, $\ub$ color, $\mbox{SFR}_{UV}$, and $\mbox{sSFR}_{UV}$. We find a monotonic trend between \textrm{C}~\textsc{iii}] EW and $\mbox{M}_{B}$ such that \textrm{C}~\textsc{iii}] EW increases at fainter values of $\mbox{M}_{B}$ and is significantly stronger (3.82$\sigma$) in the faintest $\mbox{M}_{B}$ bin.\footnote{The significance levels indicated in parentheses were determined by grouping data points into two sets according to \textrm{C}~\textsc{iii}] EW (relatively large or small), calculating their respective means and errors in \textrm{C}~\textsc{iii}] EW, and dividing the difference in their mean values by the square root of the sum in quadrature of the overall errors of the two sets of data.} A similar connection between \textrm{C}~\textsc{iii}] EW and luminosity has been observed in previous studies at $z\sim2$ \citep{Stark2014}.

The scaling relation of \textrm{C}~\textsc{iii}] strength with stellar mass is shown in the upper right panel of Figure \ref{fig:stack_ew}. Despite the \textrm{C}~\textsc{iii}] emission showing less dependence on stellar mass than on $B-$band absolute magnitude, the two bins with lower stellar masses display significantly larger (3.29$\sigma$) \textrm{C}~\textsc{iii}] EWs on average than the higher stellar mass bins. This trend could also be interpreted as a secondary correlation with the $B-$band absolute magnitude, if we assume a constant mass-to-light ratio for the galaxies in our sample.

As for $\ub$ color, although the correlation is not as evident as with the $B-$band absolute magnitude, the \textrm{C}~\textsc{iii}] EW is significantly higher (4.54$\sigma$) in the bluest bin. This finding agrees with the results from individual spectra, that significant \textrm{C}~\textsc{iii}] detections are found in bluer galaxies (Figure \ref{fig:galprop}, bottom). The reddest bin does not show the smallest \textrm{C}~\textsc{iii}] EW, contrary to the expectation based on the trend discovered in individual \textrm{C}~\textsc{iii}] detections. However, the increase in \textrm{C}~\textsc{iii}] strength in the reddest bin with respect to the middle two bins is not significant, especially given the comparatively large error bar on the corresponding \textrm{C}~\textsc{iii}] EW measurement. It is worth noting that none of the objects going into this bin have a significant individual \textrm{C}~\textsc{iii}] detections except 22036975 (Figure \ref{fig:galprop}).\footnote{The \textrm{C}~\textsc{iii}] EW remains almost unchanged (rest-frame EW decreases by 0.06$\mbox{\AA}$, which is less than half of the associated $1\sigma$ error bar) if 22036975 is removed from the reddest bin. We note that this result is a natural outcome of our approach of performing median stackings.} Furthermore, the spectra used to create the reddest $\ub$ composite were in general noisier than those in the bluer bins, especially for spectra in which \textrm{C}~\textsc{iii}] fell on the blue side. Therefore, it is the relatively poor data quality in the far-UV that is responsible for the large uncertainty associated with the reddest $\ub$ composite.

One possible origin of the enhanced \textrm{C}~\textsc{iii}] emission observed in the reddest $\ub$ bin is potential AGN contribution. [\textrm{Ne}~\textsc{v}]$\lambda$3425, a high-ionization transition in the near-UV, is one of the common signatures of AGN activity. While the LRIS spectra have [\textrm{Ne}~\textsc{v}] coverage for many objects in our sample, there are cases where this feature falls in the gap between the blue and red spectral coverage and is therefore not available. As a result, we utilized the DEEP2 spectra observed with the Deep Imaging Multi-Object Spectrograph \citep[DEIMOS;][]{Faber2003} on the Keck II telescope as a complementary dataset for objects without [\textrm{Ne}~\textsc{v}] coverage in the LRIS spectra. The DEEP2/DEIMOS spectra have a high spectral resolution (maximum $R \sim 6000$) and spectral coverage of $\sim6500-9300\mbox{\AA}$, which corresponds to the near-UV and blue optical band at the DEEP2 redshift ($z \sim 1$) and thus enables us to inspect multiple rest-optical emission features. Out of 46 objects in the reddest $\ub$ bin, 32 have [\textrm{Ne}~\textsc{v}]$\lambda$3425 coverage (3 in the LRIS spectra and 29 in the DEIMOS spectra) and none of them shows a $\geqslant3\sigma$ [\textrm{Ne}~\textsc{v}] line detection. Restricting the stack to these 32 objects with both [\textrm{Ne}~\textsc{v}] coverage and no significant [\textrm{Ne}~\textsc{v}] detection reduces the possibility of any AGN contribution to the \textrm{C}~\textsc{iii}] emission. The strength of \textrm{C}~\textsc{iii}] emission measured from this more controlled composite spectrum is shown as the gray square in the bottom panel of Figure \ref{fig:stack_ew}. In this stack, the \textrm{C}~\textsc{iii}] EW is smaller than in the original composite with all objects included, and similar to the \textrm{C}~\textsc{iii}] EW measured in the two middle $\ub$ bins. Accordingly, we conclude that there is no strong evidence for an increase in \textrm{C}~\textsc{iii}] at the reddest U-B colors.

In order to evaluate the extent to which AGN contamination affects our sample in general, we examined the DEIMOS and LRIS spectra for 164 objects with [\textrm{Ne}~\textsc{v}] coverage. None of these objects shows a $\geqslant3\sigma$ [\textrm{Ne}~\textsc{v}] line detection. We then constructed corresponding composite spectra by dividing these 164 objects into 4 bins according to $\mbox{M}_{B}$, $\mbox{M}_{*}$, and $\ub$, with each bin containing 41 objects. The rest-frame EWs of \textrm{C}~\textsc{iii}] measured from these more restricted stacks remain almost unchanged with respect to those measured from the original composite spectra, except that the \textrm{C}~\textsc{iii}] emission becomes weaker in the ``cleaned" reddest $\ub$ bin. In fact, the \textrm{C}~\textsc{iii}] EW in the ``cleaned" reddest $\ub$ bin is very similar to that in the stack made out of a set of 32 objects with [\textrm{Ne}~\textsc{v}] coverage and insignificant [\textrm{Ne}~\textsc{v}] line detections from the original reddest $\ub$ bin. Therefore, we believe that AGN contamination is a negligible effect for the majority of objects in our sample.

In summary, we observe consistent trends of \textrm{C}~\textsc{iii}] EW increasing with fainter values of $B-$band absolute magnitude, bluer $\ub$ color and smaller stellar mass in both individual and composite spectra. These results imply that strong \textrm{C}~\textsc{iii}] emitters are more likely to be faint, blue and low-mass galaxies.

\subsection{\textrm{C}~\textsc{iii}] vs. near-UV lines}
\label{sec:nuv}

Our LRIS spectra typically cover a set of several near- and far-UV emission features, in addition to \textrm{C}~\textsc{iii}]$\lambda\lambda$1907, 1909. These include \textrm{C}~\textsc{iv}$\lambda\lambda$1548, 1550, \textrm{He}~\textsc{ii}$\lambda$1640, \textrm{O}~\textsc{iii}]$\lambda\lambda$1661, 1666 and \textrm{Si}~\textsc{iii}]$\lambda\lambda$1882, 1892, the near-UV \textrm{Fe}~\textsc{ii}*$\lambda$2365, 2396, 2612, 2626 fluorescent transitions, and the \textrm{Mg}~\textsc{ii} $\lambda\lambda$2796, 2803 emission doublets. Although we only observe weak \textrm{C}~\textsc{iv}$\lambda\lambda$1548, 1550, \textrm{He}~\textsc{ii}$\lambda$1640, \textrm{O}~\textsc{iii}]$\lambda\lambda$1661, 1666 in the overall composite spectrum, strong near-UV \textrm{Fe}~\textsc{ii}* and \textrm{Mg}~\textsc{ii} transitions are well-detected in the stacked spectra described in Section \ref{sec:galprop} and numerous individual spectra.

It has been illustrated that galaxies with strong \textrm{Fe}~\textsc{ii}* and \textrm{Mg}~\textsc{ii} emission typically have higher sSFRs, lower stellar masses and bluer UV slopes \citep{Erb2012,Kornei2013}. In light of the fact that \textrm{C}~\textsc{iii}] emission displays similar correlations with stellar mass and color, it is possible that the strength of \textrm{C}~\textsc{iii}] is related to that of the \textrm{Fe}~\textsc{ii}* and \textrm{Mg}~\textsc{ii} emission. In order to investigate the relationship between \textrm{C}~\textsc{iii}] and the near-UV transitions and potentially study the possible underlying physical connection between the gas components these lines trace, we measured both the \textrm{C}~\textsc{iii}] and the near-UV \textrm{Fe}~\textsc{ii}* and \textrm{Mg}~\textsc{ii} lines in the stacked spectra. For this analysis, we used the $\mbox{M}_{B}$, $\ub$, $\mbox{SFR}_{UV}$, and $\mbox{sSFR}_{UV}$ stacks described in Section \ref{sec:galprop}, where the $\mbox{M}_{B}$ and $\ub$ stacks were based on the full sample of objects with \textrm{C}~\textsc{iii}] coverage, and the $\mbox{SFR}_{UV}$ and $\mbox{sSFR}_{UV}$ stacks only included 42 objects with $\mbox{SFR}_{UV}$ measurements .

\begin{figure}
\includegraphics[width=1.0\linewidth]{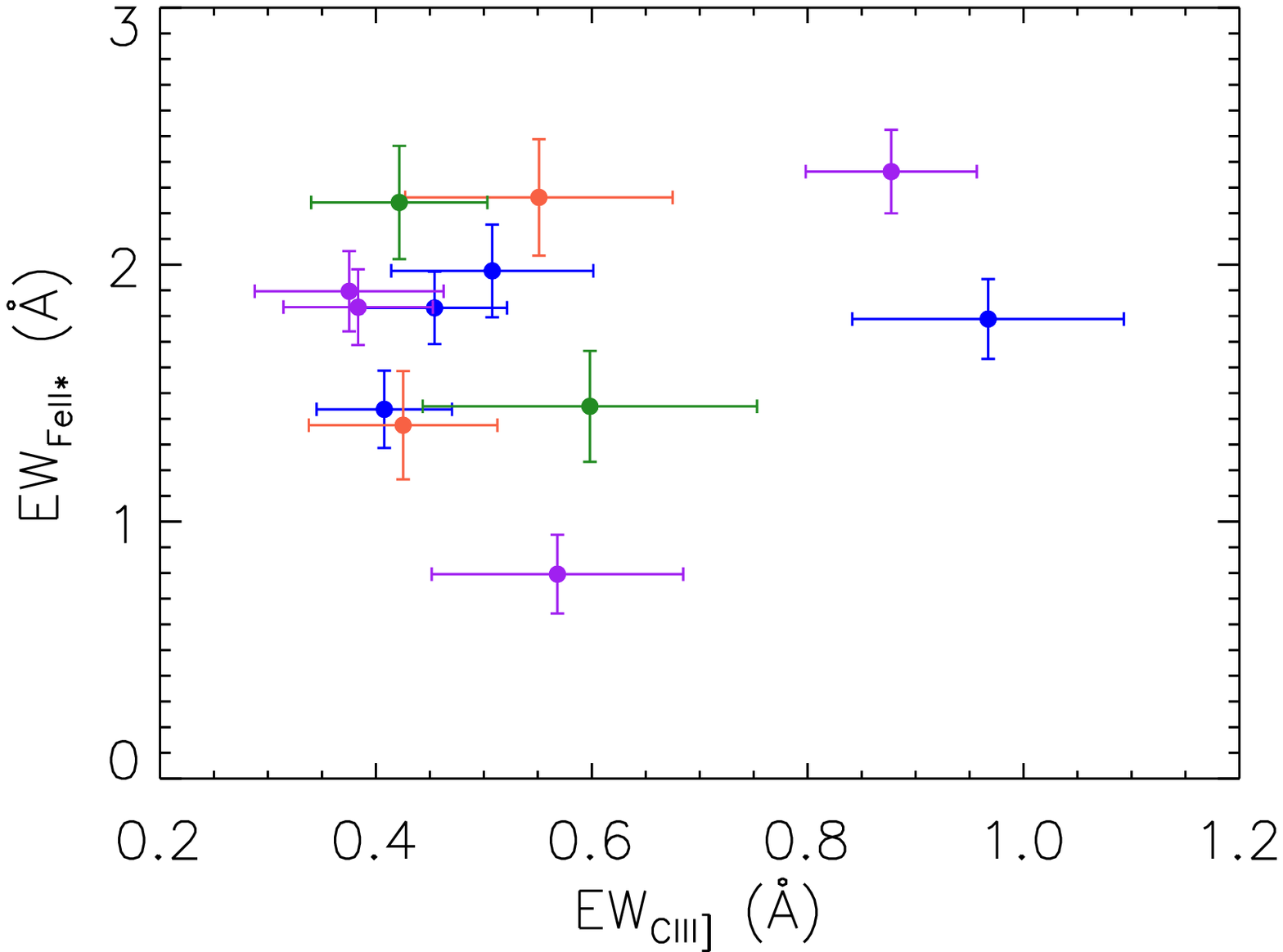}
\includegraphics[width=1.0\linewidth]{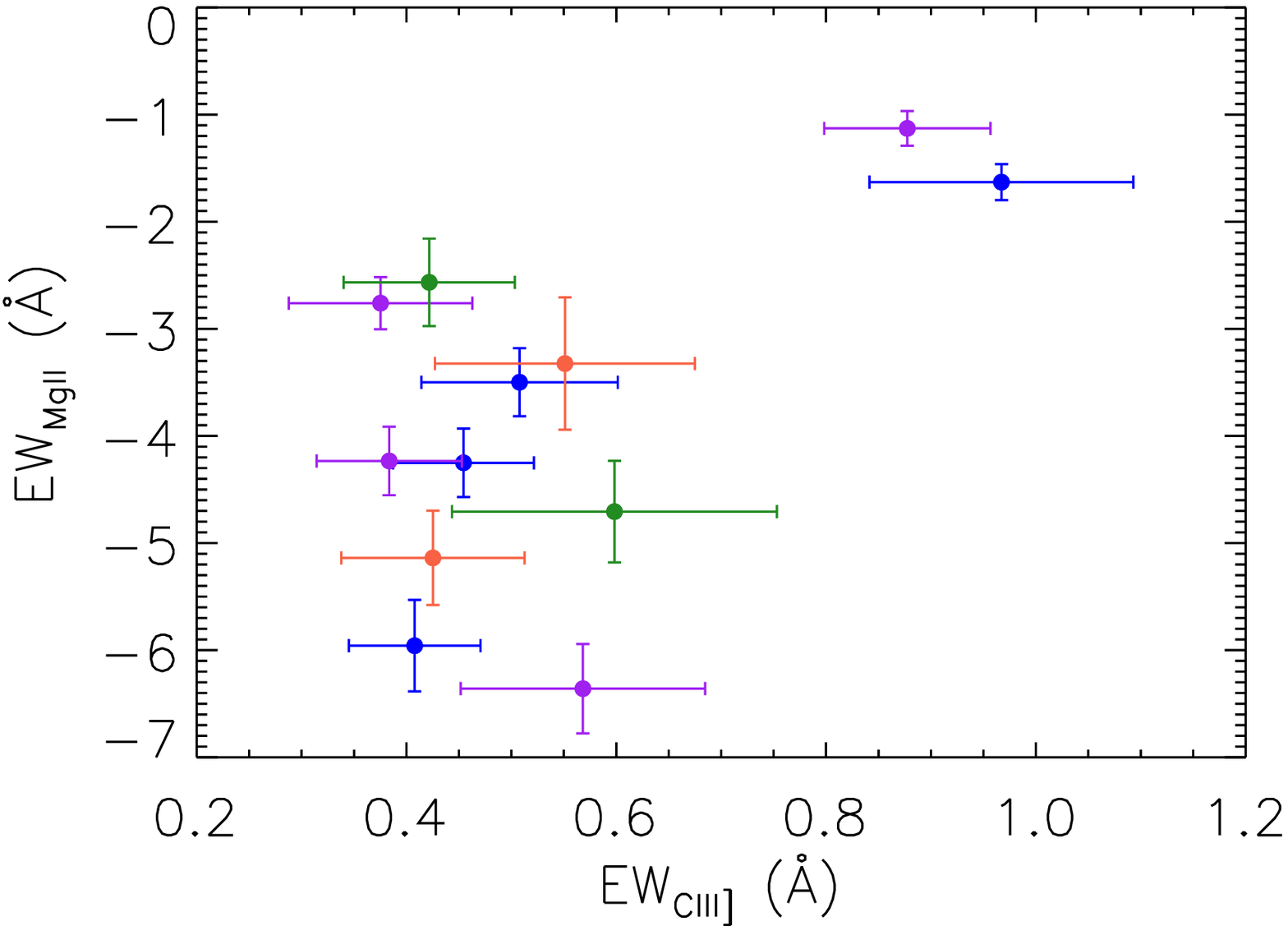}
\caption{\textbf{Top:} \textrm{C}~\textsc{iii}] EW vs. near-UV \textrm{Fe}~\textsc{ii}* EW in the composite spectra described in Section \ref{sec:galprop}. The combined EW of \textrm{Fe}~\textsc{ii}*$\lambda$2365, 2396, 2612, 2626 is shown. Blue, purple, red and green circles correspond to measurements from stacked spectra of $\mbox{M}_{B}$, $\ub$ color, $\mbox{SFR}_{UV}$ and $\mbox{sSFR}_{UV}$, respectively. \textbf{Bottom:} \textrm{C}~\textsc{iii}] EW vs. \textrm{Mg}~\textsc{ii} $\lambda\lambda$2796, 2803 absorption EW in the composite spectra.}
\label{fig:nuv}
\end{figure}

To obtain a characteristic \textrm{Fe}~\textsc{ii}* line strength, we fit \textrm{Fe}~\textsc{ii}*$\lambda$2365, 2396, 2612 and 2626 individually with single Gaussian profiles using MPFIT and combined the EWs of those 4 lines. The case for \textrm{Mg}~\textsc{ii} is more complicated. It is difficult to obtain a robust estimate of the separate contribution of \textrm{Mg}~\textsc{ii} emission, as the \textrm{Mg}~\textsc{ii} emission and absorption components are significantly blended. The presence of emission is indirectly inferred from the apparent \textrm{Mg}~\textsc{ii} absorption profile, which has a more blueshifted centroid (due to the filling in from \textrm{Mg}~\textsc{ii} emission on the red side), and a greater variation in strength than the near-UV \textrm{Fe}~\textsc{ii} absorption lines \citep{Martin2012, Kornei2013}. Therefore, we used the variation of the \textrm{Mg}~\textsc{ii}$\lambda\lambda$2796, 2803 absorption doublet as a proxy for the associated \textrm{Mg}~\textsc{ii} emission, and examined the correlation between apparent \textrm{Mg}~\textsc{ii} absorption EW and \textrm{C}~\textsc{iii}] emission. Figure \ref{fig:nuv} shows the correlations between the strength of \textrm{C}~\textsc{iii}] and that of \textrm{Fe}~\textsc{ii}* emission (top) and \textrm{Mg}~\textsc{ii} absorption (bottom).

We find that the \textrm{Fe}~\textsc{ii}* strength does not change significantly (2.89$\sigma$) with \textrm{C}~\textsc{iii}] EW, while \textrm{C}~\textsc{iii}] emission is noticeably stronger (16.6$\sigma$) in the two composite spectra with the weakest \textrm{Mg}~\textsc{ii} absorption. Furthermore, the decline in the absorption strength is much more pronounced in \textrm{Mg}~\textsc{ii} than in the near-UV \textrm{Fe}~\textsc{ii} in those two composites. As the \textrm{Fe}~\textsc{ii} absorption profiles are less affected by resonance emission filling \citep{Erb2012}, they can be used as a proxy for the intrinsic, unaffected profiles of \textrm{Mg}~\textsc{ii} absorption. Given the greater EW decrease in \textrm{Mg}~\textsc{ii} absorption than \textrm{Fe}~\textsc{ii} absorption (i.e., where the \textrm{Fe}~\textsc{ii} variation indicates the change in underlying absorption), we interpret the observed trend between \textrm{C}~\textsc{iii}] and \textrm{Mg}~\textsc{ii} as a joint effect of \textrm{Mg}~\textsc{ii} absorption becoming weaker and \textrm{Mg}~\textsc{ii} emission becoming stronger with increasing \textrm{C}~\textsc{iii}] emission EW \citep{Martin2012,Kornei2013}. The distinct behavior between \textrm{Mg}~\textsc{ii} and \textrm{Fe}~\textsc{ii}* is likely a result of the difference in the underlying source function. Although both \textrm{Mg}~\textsc{ii} and \textrm{Fe}~\textsc{ii}* photons are scattered by circumgalactic gas \citep{Rubin2011,Martin2013}, the source function for \textrm{Mg}~\textsc{ii} includes an emission line (i.e., nebular emission from the \textrm{H}~\textsc{ii} regions), which becomes stronger in lower-mass galaxies. The source function for \textrm{Fe}~\textsc{ii}*, however, is a flat continuum \citep{Erb2012}, which is absorbed and re-emitted in the possibly outflowing circumgalactic gas \citep{Erb2012, Kornei2013}, resulting in \textrm{Fe}~\textsc{ii} absorption and \textrm{Fe}~\textsc{ii}* emission transitions, respectively.

\subsection{\textrm{C}~\textsc{iii}] vs. Metallicities}
\label{sec:metal}

Photoionization models and observations at different redshifts have revealed that low gas metallicity leads to enhanced nebular \textrm{C}~\textsc{iii}] emission due to the hard ionizing SED of the associated metal-poor stars exciting the gas, and the elevated electron temperature resulting from reduced metal cooling \citep[e.g., ][see Section \ref{sec:photo} for a detailed discussion]{Erb2010, Stark2014, Rigby2015,Jaskot2016}. In order to examine if the correspondence between \textrm{C}~\textsc{iii}] emission and metallicity found in high-redshift studies also holds for $z \sim 1$, we estimated metallicities using optical emission features.

Ideally, we would like to use the combination of [\textrm{O}~\textsc{ii}]$\lambda3727$, [\textrm{O}~\textsc{iii}]$\lambda5007$ and $\mbox{H}\beta$ to obtain estimates of both metallicity and ionization parameter (e.g., $R_{23}\equiv([\textrm{O}~\textsc{ii}]+[\textrm{O}~\textsc{iii}])/\mbox{H}\beta$ and $O_{32}\equiv[\textrm{O}~\textsc{iii}]/[\textrm{O}~\textsc{ii}]$). Since [\textrm{O}~\textsc{iii}] and $\mbox{H}\beta$ are not typically covered in the LRIS spectra, we inspected the DEIMOS spectra for 26 objects in the DEEP2/LRIS sample with redshift lower than $\sim$ 0.9, where these optical diagnostic features fall within the DEIMOS spectral coverage. However, the lack of robust flux calibration and nebular dust extinction estimate for the DEIMOS spectra led us to focus instead on the simple combination of [\textrm{O}~\textsc{iii}] and $\mbox{H}\beta$, which are relatively close in wavelength. The measured [\textrm{O}~\textsc{iii}]/$\mbox{H}\beta$ flux ratio should not be significantly affected by dust extinction, and, due to the proximity of the lines, this flux ratio can be estimated even if the spectrum does not have a precise flux calibration.

[\textrm{O}~\textsc{iii}]/$\mbox{H}\beta$ shows a dependence on the gas metallicity, $12+\log(\mbox{O}/\mbox{H})$, such that at very low metallicities [\textrm{O}~\textsc{iii}]/$\mbox{H}\beta$ increases with increasing $12+\log(\mbox{O}/\mbox{H}$), as a result of increasing oxygen abundance. [\textrm{O}~\textsc{iii}]/$\mbox{H}\beta$ reaches a maximum value at $12+\log(\mbox{O}/\mbox{H})\sim 8.0$, and then declines with increasing metallicity due to lower electron temperature and lower ionization state \citep{Maiolino2008, Jones2015}. Given the form of this relationship, the translation between  [\textrm{O}~\textsc{iii}]/$\mbox{H}\beta$ is double valued over a wide range in [\textrm{O}~\textsc{iii}]/$\mbox{H}\beta$. However, it is possible to use external information to determine on which branch of the [\textrm{O}~\textsc{iii}]/$\mbox{H}\beta$ vs. metallicity relation a given galaxy falls. For example, given the mass-metallicity relationship (MZR) observed among star-forming galaxies over a wide range of redshifts \citep{Tremonti2004,Erb2006,Maiolino2008}, and the stellar mass range of the 26 objects in our sample with $z<0.9$ ($\log(\mbox{M}_{*}/\mbox{M}_{\sun}) = 8.88 -11.21$), we infer that these objects fall within the high stellar-mass, high-metallicity branch, despite the uncertainties associated with the MZR \citep[0.1 dex, ][]{Tremonti2004} and the [\textrm{O}~\textsc{iii}]/$\mbox{H}\beta$ vs. metallicity relation \citep[$\sim0.1$ dex, ][]{Maiolino2008}. In this regime,  [\textrm{O}~\textsc{iii}]/$\mbox{H}\beta$ monotonically decreases with increasing metallicity.

\begin{figure}
\includegraphics[width=1.0\linewidth]{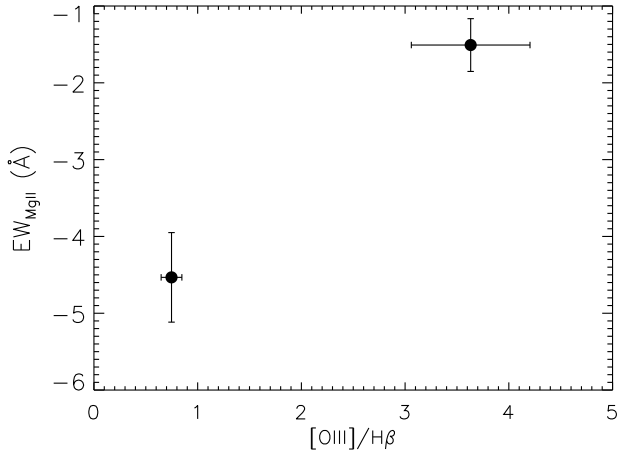}
\caption{EW of \textrm{Mg}~\textsc{ii} $\lambda\lambda$2796, 2803 absorption doublet vs. [\textrm{O}~\textsc{iii}]/$\mbox{H}\beta$ in the composite spectra. There are 26 objects at $z<0.9$ in the DEEP2/LRIS sample with [\textrm{O}~\textsc{iii}] and $\mbox{H}\beta$ coverage in their DEIMOS spectra. These objects were used to generate the composite spectra.}
\label{fig:o3hb}
\end{figure}

We fit the [\textrm{O}~\textsc{iii}] and $\mbox{H}\beta$ emission profiles with single Gaussians using MPFIT, the same procedure as described in Section \ref{sec:meas}. As we do not detect individual \textrm{C}~\textsc{iii}] emission in any of these 26 objects, we performed binary stacks according to the [\textrm{O}~\textsc{iii}]/$\mbox{H}\beta$ ratio, with the low and high bins containing nearly equal numbers of objects. Although the \textrm{C}~\textsc{iii}] feature in both stacked spectra is not significant, we find that \textrm{Mg}~\textsc{ii} absorption becomes weaker (or equivalently, \textrm{Mg}~\textsc{ii} emission becomes stronger) with higher [\textrm{O}~\textsc{iii}]/$\mbox{H}\beta$. Based on the correlation in strength between \textrm{C}~\textsc{iii}] and \textrm{Mg}~\textsc{ii} emission (Section \ref{sec:nuv}), this result suggests indirectly that stronger \textrm{C}~\textsc{iii}] may be found within lower-metallicity galaxies in our sample. Our inferred correlation between \textrm{C}~\textsc{iii}] and metallicity is not only consistent with both models and observations of \textrm{C}~\textsc{iii}] at other redshifts \citep{Stark2014, Rigby2015,Jaskot2016}, but is also supported by the result that CIII] emission is significantly stronger in the lower-mass half of our sample, given the well-measured correlation between metallicity and stellar mass \citep[e.g., ][]{Tremonti2004,Zahid2013}.

\section{Discussion}
\label{sec:dis}

\subsection{\textrm{C}~\textsc{iii}] EW at other redshifts}
\label{sec:ew_z}

\begin{figure*}
\includegraphics[width=1.0\linewidth]{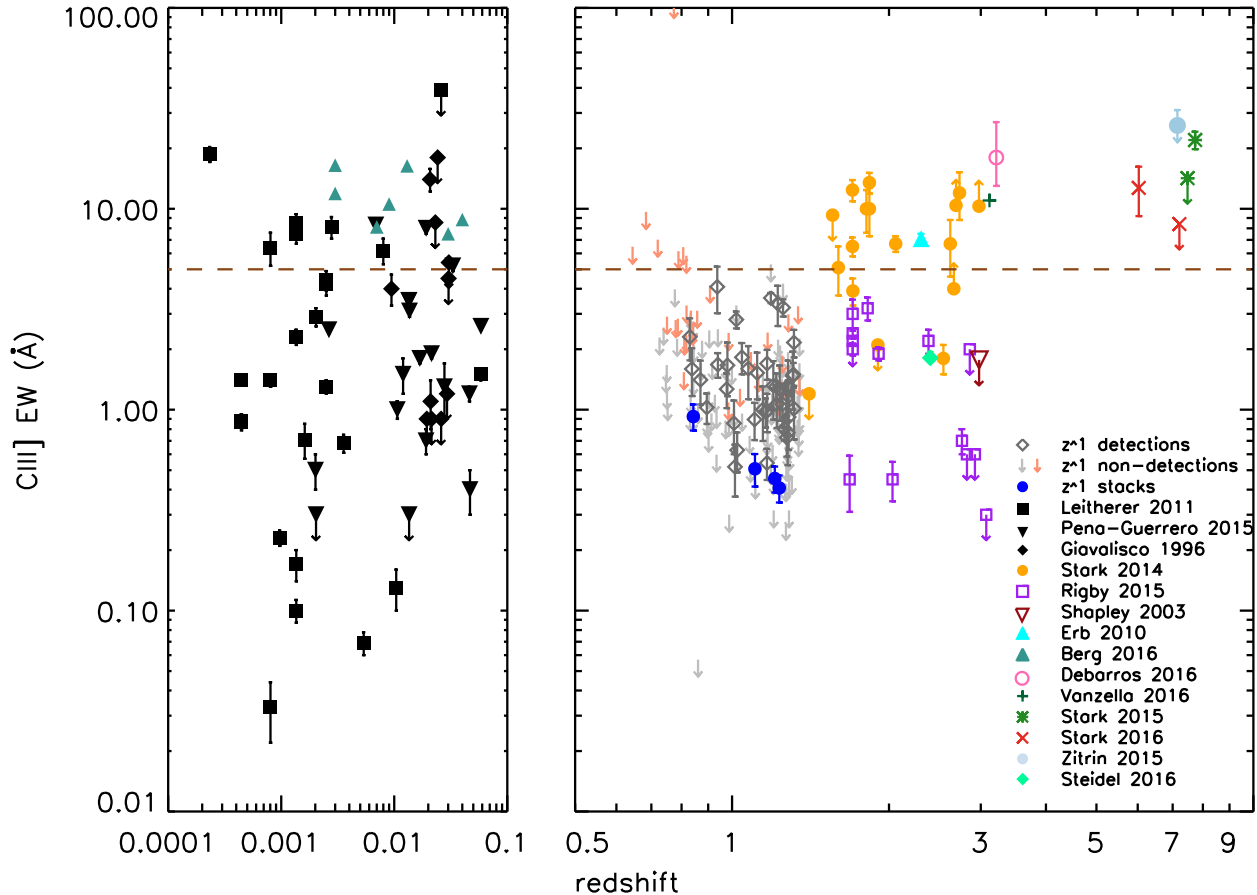}
\caption{\textrm{C}~\textsc{iii}] EW at different redshifts, with measurements drawn from the literature. The error bars represent $1\sigma$ uncertainties for significant detections, while the downward-pointing arrows indicate $3\sigma$ upper limits for EW measurements that are not significant. Measurements from the individual \textrm{C}~\textsc{iii}] detections and composite spectra in 4 $\mbox{M}_{B}$ bins from the DEEP2/LRIS sample are shown, respectively, in gray open diamonds and dark blue circles. The gray and pink arrows indicate the $z\sim1$ non-detections with continuum $S/N>5$ and $S/N<5$, respectively. The local samples include measurements from \citet[][black diamonds]{Gia1996}, \citet[][black circles]{Leitherer2011}, Pena-Guerrero et al. (in prep.) (black triangles) and \citet[][dark cyan triangles]{Berg2016}. The high-redshift samples include measurements from \citet[][orange circles]{Stark2014}, \citet[][purple squares]{Rigby2015}, \citet[][light green diamond]{Steidel2016}, \citet[][dark red triangle]{Shapley2003}, \citet[][cyan triangle]{Erb2010}, \citet[][green plus sign]{Vanzella2016}, \citet[][pink circle]{Debarros2016}, \citet[][red crosses]{Stark2016}, \citet[][green stars]{Stark2015}, and \citet[][light blue circles]{Zitrin2015}. The brown dashed line marks where EW$=5\mbox{\AA}$.}
\label{fig:z_ew}
\end{figure*}

As a promising alternative redshift diagnostic, \textrm{C}~\textsc{iii}] has been investigated in various types of galaxies in multiple high-redshift ($2 \lesssim z \lesssim 7$) studies of UV emission lines. Large \textrm{C}~\textsc{iii}] EWs ($\geqslant5\mbox{\AA}$, as defined in Rigby et al. 2015) have been identified in galaxies at high redshift ($z\geqslant2$) and in the local universe. Several studies have measured \textrm{C}~\textsc{iii}] EWs $\gtrsim10\mbox{\AA}$ in galaxies formed prior to and during the epoch of reionization ($z>6$). \citet{Stark2015} studied two star-forming galaxies at $z\simeq6$ with Ly$\alpha$ detections, where the EWs of the \textrm{C}~\textsc{iii}] doublet were measured to be $12.7\pm3.5$ and $7.8\pm2.8\mbox{\AA}$, respectively. In addition, \citet{Zitrin2015} presented a sample of 14 lensed galaxies at $z_{phot}\simeq7-8$. Although these authors did not have individual convincing detections of \textrm{C}~\textsc{iii}] in their sample, an upper limit on the \textrm{C}~\textsc{iii}] doublet rest-frame EW was estimated to be $26\pm5\mbox{\AA}$ with $95\%$ confidence. More recently, \citet{Stark2016} observed 3 galaxies at $z\sim7$ with extreme rest-optical emission line ([\textrm{O}~\textsc{iii}]+$\mbox{H}\beta$) fluxes. These authors measured a \textrm{C}~\textsc{iii}] EW of $22\pm2\mbox{\AA}$ for EGS-zs8-1 and a $3\sigma$ upper limit of 14$\mbox{\AA}$ for EGS-zs8-2. We note that all three galaxies at $z>6$ with significant \textrm{C}~\textsc{iii}] detections also have strong Ly$\alpha$ emission (rest-frame EW$>20\AA$).

Besides observations at the high-redshift frontier, various studies have been conducted to examine the properties of galaxies at $z\sim2.5$, near the peak of the SFR density in the universe \citep[$z\sim2.5$;][]{Madau2014}. \citet{Stark2014} examined 17 $z\sim2$ low-mass, low-luminosity lensed galaxies, among which 16 have detections of \textrm{C}~\textsc{iii}]. The measured \textrm{C}~\textsc{iii}] EW has a mean of 7.1$\mbox{\AA}$ and a maximum of 13.5$\mbox{\AA}$. \textrm{C}~\textsc{iii}] has also been detected in multiple individual galaxies at similar redshifts. These include a \textrm{C}~\textsc{iii}] EW measurement of $7.1\pm0.4\mbox{\AA}$ from a low-mass, low-metallicity, unreddened system, Q2343-BX418 at $z=2.3$ \citep{Erb2010}; strong \textrm{C}~\textsc{iii}] emission ($11\mbox{\AA}$) in a low-mass, compact, star-forming galaxy at $z=3.12$ \citep{Vanzella2016}; and an $18^{+9}_{-5}\mbox{\AA}$ \textrm{C}~\textsc{iii}] detection in a low-metallicity, Lyman continuum emitter at $z=3.2$ \citep{Debarros2016}.

Observations of local galaxies have shown that strong \textrm{C}~\textsc{iii}] emission is not exclusive to the high-redshift universe. \citet{Rigby2015} presented a compilation of 46 \textrm{C}~\textsc{iii}] measurements from nearby starburst galaxies, which were analyzed in \citet{Gia1996}, \citet{Leitherer2011}, and Pena-Guerrero et al. (in prep.). The local samples show a large scatter in \textrm{C}~\textsc{iii}] EW, and $22\%$ of them show strong ($>5\mbox{\AA}$) \textrm{C}~\textsc{iii}] emission. Most of these local strong \textrm{C}~\textsc{iii}] emitters are low-metallicity galaxies, with a typical oxygen abundance $12+\log(\mbox{O}/\mbox{H})\lesssim8.0$. Moreover, \citet{Berg2016} recently reported 7 significant \textrm{C}~\textsc{iii}] detections from a sample of 12 nearby, low-metallicity ($12+\log(\mbox{O}/\mbox{H})=7.3-8.2$) dwarf galaxies with typical stellar masses $\log(\mbox{M}_{*}/\mbox{M}_{\sun})\sim6-7$. These \textrm{C}~\textsc{iii}] measurements yield a larger EW on average compared to the other local samples, with a median of $10.6\mbox{\AA}$.

While galaxies with large ($>5\mbox{\AA}$) \textrm{C}~\textsc{iii}] EW have been identified at both low and high redshift, at least at $z\leqslant3$, these systems are not common given the typical ranges in stellar mass and metallicity probed for ``main sequence" \citep{Noeske2007} galaxies. The strong \textrm{C}~\textsc{iii}] emitters tend to have low mass, luminosity (in some cases accessible due to the effects of strong gravitational lensing), and metallicity. In contrast to what is observed in these extreme systems, studies focusing on the general star-forming populations find a characteristic \textrm{C}~\textsc{iii}] EW of only $\sim1-2\mbox{\AA}$. \citet{Shapley2003} analyzed a composite spectrum created out of 811 individual Lyman Break Galaxy (LBG) spectra at $z\sim3$, and reported a weak \textrm{C}~\textsc{iii}] emission feature of $1.67\pm0.59\mbox{\AA}$ (i.e., $\sim2.8\sigma$). \citet{Rigby2015} measured \textrm{C}~\textsc{iii}] in 11 lensed star-forming galaxies at $z\sim1.6-3$, finding the median EW in their sample to be $2.0\mbox{\AA}$. More recently, \citet{Steidel2016} observed a sample 30 star-forming galaxies at $z\sim2.4$, and the estimated \textrm{C}~\textsc{iii}] EW from the composite spectrum was $1.81\pm0.13\mbox{\AA}$ (C. Steidel, private communication). \footnote{In \citet{Steidel2016}, these authors only presented the flux measurement of \textrm{C}~\textsc{iii}] from the composite spectrum. The quoted value of the rest-frame \textrm{C}~\textsc{iii}] EW resulted from our measurement from the continuum-normalized \citet{Steidel2016} composite, following the method described in Section \ref{sec:meas}. Due to the lack of the corresponding composite error spectrum, we estimated the continuum noise from a relatively featureless range ($2000-2200\mbox{\AA}$) and assigned that value as a constant error across the full spectrum. We note that in this manner, the uncertainty on the EW is underestimated, as the sample variance has not been accounted for.}

In Figure \ref{fig:z_ew} we compare our $z\sim1$ DEEP2/LRIS sample with studies discussed above in terms of \textrm{C}~\textsc{iii}] EW. As shown by the dividing line at $\mbox{EW}_{\textrm{C}~\textsc{iii}]}=5\mbox{\AA}$, none of the individual or composite spectra in our sample displays the large EW ($>5\mbox{\AA}$) observed in the extreme systems (i.e., $z>6$ galaxies, lensed $z\sim2$ galaxies, some local starbursts). Specifically, the typical EW observed in our $z\sim1$ DEEP2/LRIS sample with \textrm{C}~\textsc{iii}] detections is $\sim1\mbox{\AA}$, which is very similar to those measured in the global galaxy studies at $z\sim2-3$ \citep[e.g., ][]{Shapley2003, Steidel2016}. This result suggests that typical star-forming galaxies at $z\sim1-3$ (with stellar masses of $\log(\mbox{M}_{*}/\mbox{M}_{\sun})\sim9-11.5$) are quite distinct in rest-UV emission properties from the galaxies studied thus far at $z>6$.

\subsection{Physical Conditions at $z\sim1$}
\label{sec:phy}

As discussed in Section \ref{sec:ew_z}, we observe a small typical \textrm{C}~\textsc{iii}] EW ($\sim1\mbox{\AA}$) at $z\sim1$ and detect no strong \textrm{C}~\textsc{iii}] emitters ($>5\mbox{\AA}$) like those discovered at both higher and lower redshift. In order to gain a comprehensive understanding of the physical conditions in \textrm{H}~\textsc{ii} regions in star-forming galaxies at $z\sim1$, we present estimates of the gas-phase metallicity and C/O abundance for a small subset of objects in our sample, and compare our results with the predictions from photoionization models.

\subsubsection{Gas Metallicity and C/O Ratio}
\label{sec:co}

In Section \ref{sec:metal}, we have used [\textrm{O}~\textsc{iii}]$\lambda$5007/H$\beta$ as an indicator of gas metallicity for a subset of 26 galaxies in our $z\sim1$ sample. The median [\textrm{O}~\textsc{iii}]/$\mbox{H}\beta$ measured in the 26 DEIMOS spectra is 2.50, which corresponds to a gas metallicity of $12+\log(\mbox{O}/\mbox{H})\sim8.4$ \citep[i.e., $\sim0.5\mbox{Z}_{\sun}$;][]{Maiolino2008,Jones2015}. Our metallicity measurement is $\sim$0.5 dex lower at ${10}^{10} \mbox{M}_{\sun}$ (close to the median stellar mass of our sample) than that in \citet{Zahid2013} estimated from $z\sim0.8$ DEEP2 galaxies. For this comparison, we use the same metallicity indicator ([\textrm{O}~\textsc{iii}]/$\mbox{H}\beta$), and calibration \citep{Maiolino2008} for both samples. This discrepancy in metallicity primarily comes from sample selection.  Galaxies in our DEEP2/LRIS sample are predominantly ``blue cloud'' star-forming galaxies and are bluer in $\ub$ color on average than the full DEEP2 sample \citep{Martin2012}. Based on the assumption that $\ub$ color tracks sSFR, and that the fundamental metallicity relationship \citep[e.g., ][]{Mannucci2010} holds at $z\sim1$, the $\mbox{O}/\mbox{H}$ ([\textrm{O}~\textsc{iii}]/$\mbox{H}\beta$) in our sample should be systematically lower (higher) than in the full DEEP2 sample from \citet{Zahid2013}. We also note that given the small number of objects (26 galaxies) with [\textrm{O}~\textsc{iii}]/$\mbox{H}\beta$ measurements, the quoted median [\textrm{O}~\textsc{iii}]/$\mbox{H}\beta$ value may not be representative of the full DEEP2/LRIS sample.

The C/O ratio has been found to correlate with oxygen abundance, which is due to either the weaker stellar winds in metal-poor stars, or the delay in carbon enrichment in the ISM from low-mass stars \citep[e.g.,][]{Henry2000,Akerman2004,Erb2010}. Carbon is on average more abundant in high-metallicity environments \citep[e.g.,][]{Erb2010,Berg2016}. However, the softer radiation field and lower electron temperature (due to enhanced metal cooling) associated with higher oxygen abundance result in reduced \textrm{C}~\textsc{iii}] strength, with the net effect being weaker \textrm{C}~\textsc{iii}] emission at higher C/O ratio.

Given the spectral coverage of the LRIS spectra, we were able to additionally constrain the C/O ratio, which can be measured from the observed flux ratio of [\textrm{O}~\textsc{iii}]$\lambda\lambda$1661, 1666 and \textrm{C}~\textsc{iii}]$\lambda$1909. We created a composite spectrum out of 128 objects with both [\textrm{O}~\textsc{iii}] and \textrm{C}~\textsc{iii}] coverage in the LRIS data. In order to scale these spectra to the same flux level, we took the median flux value of individual 1-D, flux-calibrated spectra over the wavelength range of $2110-2200\mbox{\AA}$ as the characteristic value for corresponding objects. We then scaled the spectra accordingly to the median of those 128 ``characteristic values." The 600-line spectra were smoothed to the resolution of the 400-line spectra, and the median value at each pixel was extracted to generate the composite spectrum. Due to noise and the intrinsic weak nature of the [\textrm{O}~\textsc{iii}]$\lambda$1661 transition, we did not have a robust detection of that feature. Instead, we adopted the doublet ratio (i.e., $\mbox{I}_{[\textrm{O}~\textsc{iii}]\lambda1666}/\mbox{I}_{[\textrm{O}~\textsc{iii}]\lambda1661}$) of 2.48 for the doublet members, and inferred the strength of [\textrm{O}~\textsc{iii}]$\lambda$1661 from the stronger doublet member, [\textrm{O}~\textsc{iii}]$\lambda1666$. We then fit [\textrm{O}~\textsc{iii}]$\lambda1666$ and \textrm{C}~\textsc{iii}]$\lambda$1909 separately with local continuum using MPFIT. 

The best-fit parameters yield a \textrm{C}~\textsc{iii}]/[\textrm{O}~\textsc{iii}] flux ratio of 2.58$\pm$0.57, which corresponds to a $\mbox{C}^{++}$/$\mbox{O}^{++}$ abundance ratio of 0.13$\pm$0.03 following Equation 3 in \citet{Erb2010}, assuming a typical temperature of ${10}^{4}$ K for \textrm{H}~\textsc{ii} regions in our sample galaxies, and neglecting the effect of dust extinction. We approximated this value as the C/O abundance ratio with the assumption of a unity ionization correction factor (ICF), which results in $\log(C/O)=-0.89^{+0.09}_{-0.11}$ (i.e., $\sim0.3\mbox{(C/O)}_{\sun}$). Compared with the C/O vs. metallicity relation observed in local galaxies and \textrm{H}~\textsc{ii} regions \citep{Garnett1995,Erb2010,Berg2016}, the C/O ratio we measured is $\sim0.4$ dex lower at a metallicity of $12+\log(\mbox{O}/\mbox{H})\sim8.4$ ($\sim0.5 \mbox{Z}_{\sun}$). This low C/O value could possibly result from the assumption of ICF$=1$, which implies that $\mbox{C}^{++}$ and $\mbox{O}^{++}$ characterize the overall carbon and oxygen abundances. However, if the ICF in our $z\sim1$ galaxies is larger than unity due to, for example, larger ionization parameters \citep[$\log\mbox{U}\geqslant-2$;][]{Erb2010,Gutkin2016}, the inferred C/O ratio would increase and become consistent with other local measurements. Furthermore, we note that the formal error bar on the C/O flux ratio estimated from MPFIT does not include the systematic uncertainty in the placement of the continuum level, which could potentially affect the flux measurement of weak features such as \textrm{O}~\textsc{iii}]$\lambda\lambda$1661, 1666.

\subsubsection{Photoionization Models}
\label{sec:photo}

The correlations between \textrm{C}~\textsc{iii}] strength and galaxy properties have been investigated not only from the observational side, but also from a theoretical perspective. Multiple studies have used photoionization models to examine the factors that regulate the emission profile of \textrm{C}~\textsc{iii}]. \citet{Stark2014} combined a stellar population synthesis model with a photoionization code to describe the ensemble of \textrm{H}~\textsc{ii} regions and the diffuse gas ionized by young stars. These authors concluded that metal-poor gas ($\lesssim0.2-0.4\mbox{Z}_{\sun}$) and stars, young stellar populations ($6-50$ Myr), subsolar C/O ratios and large ionization parameters were responsible for large \textrm{C}~\textsc{iii}] EWs. More recently, \citet{Jaskot2016} analyzed the \textrm{C}~\textsc{iii}] EW and various emission line ratios as a function of multiple properties (e.g., starburst age, metallicity and ionization parameter) in star-forming galaxies using Cloudy\citep{Ferland1998} photoionization models. In these models, \citet{Jaskot2016} incorporated a range of C/O abundance, dust content, gas density, optical depth, and different nebular geometries. These authors find that low metallicities and high ionization parameters enhance \textrm{C}~\textsc{iii}] emission, and the highest \textrm{C}~\textsc{iii}] EWs are observed in models including binary interactions among the massive stars whose radiation ionizes the \textrm{H}~\textsc{ii} regions \citep[e.g., ][]{Stanway2016}. Furthermore, \textrm{C}~\textsc{iii}] emission peaks at young galaxy ages ($\lesssim3$ Myr after an instantaneous burst or $\lesssim10$ Myr of continuous star formation), lower dust content, and higher nebular densities. Shell-like geometries and shocks are also claimed as possible secondary effects to boost \textrm{C}~\textsc{iii}] emission.

These simulations lead to a physical picture in which strong \textrm{C}~\textsc{iii}] emission is a consequence of various factors. Young, metal-poor stellar populations generate a hard radiation field, ionizing the interstellar medium and elevating the electron temperature in the \textrm{H}~\textsc{ii} regions \citep{Stark2014}. With large sSFRs, more ionizing photons are produced per unit mass, increasing the density of free electrons and giving rise to large EWs in collisionally excited emission lines \citep{Stark2014}. The presence of high-energy, ionizing photons also results in a high ionization state of carbon (i.e., being in the state of $\mbox{C}^{++}$ instead of $\mbox{C}^{+}$). At the same time, the low gas metallicity prevents efficient metal cooling, further maintaining the electron temperature and securing a stable production of the nebular \textrm{C}~\textsc{iii}] emission \citep{Rigby2015}.

The absence of strong \textrm{C}~\textsc{iii}] emitters in our sample indicates that the conditions in the $z\sim1$ star-forming galaxies are significantly different from what we just described. \citet{Jaskot2016} predicted a weaker \textrm{C}~\textsc{iii}] emission of only a few $\mbox{\AA}$ at high metallicities and older ages in all stellar population models they explored (Padova, Geneva and BPASS) with either instantaneous or continuous star-formation regardless of the ionization parameter ($\log\mbox{U}$ from -1 to -4) and hydrogen density ($\log\mbox{n}_{H}$ from 1 to 4). Specifically, the \textrm{C}~\textsc{iii}] EW declines to $\lesssim1\mbox{\AA}$, the typical value we observe in our $z\sim1$ sample, in several cases with continuous star formation\footnote{We consider the continuous star-formation model as a better description of the star-forming history than the instantaneous burst model for majority of the galaxies in our sample. Therefore, although low \textrm{C}~\textsc{iii}] EW ($\lesssim1\mbox{\AA}$) has also been predicted in a few instantaneous burst models, we do not include those results in our discussion here.}: all models with $\mbox{Z}\geqslant0.008$ (0.57 $\mbox{Z}_{\sun}$) at ages $>20$ $\mbox{Myr}$ with hydrogen density $\log\mbox{n}_{H}\leqslant3$, and models with lower metallicities ($\mbox{Z}\leqslant0.004$ or 0.29 $\mbox{Z}_{\sun}$) at ages $>20$ $\mbox{Myr}$ with moderate to low ionization parameter and hydrogen density ($\log\mbox{U}=-3$ - $-4$), $\log\mbox{n}_{H}=1-3$.\footnote{The BPASS models, which take into account the binary interactions and rotation, produce higher \textrm{C}~\textsc{iii}] EW than the Padova and Geneva models with the same input parameters.} Combining the outputs of photoionization models and our estimates of metallicity and C/O ratio, we infer that the typical $z\sim1$ star-forming galaxies in the stellar mass range probed here have high gas-phase metallicities ($\gtrsim0.5\mbox{Z}_{\sun}$), subsolar C/O ratios, and stellar population ages $>50$ Myr.\footnote{In fact, we estimated the typical ages of our sample galaxies to be several hundreds of Myr based on their stellar masses and specific SFRs \citep{Martin2012}.}

Spectra with high $S/N$ and spectral resolution are required for the detection of weak far-UV emission lines (i.e., \textrm{C}~\textsc{iii}]$\lambda\lambda$1907, 1909, \textrm{C}~\textsc{iv}$\lambda\lambda$1548, 1550, [\textrm{O}~\textsc{iii}]$\lambda\lambda$1661, 1666). Such data will enable robust measurements of multiple physical parameters, such as ionization parameter and C/O abundance for individual objects, and provide a comprehensive picture of the physical environment in typical star-forming galaxies at $z\sim1$.

\section{Summary}
\label{sec:sum}

\textrm{C}~\textsc{iii}] has been proven to be a key emission feature in the far-UV, not only because it can potentially be used as an alternative redshift indicator to Ly$\alpha$ for galaxies at $z>6$, but also because of the useful information it provides regarding the physical conditions of star-forming regions. In this paper, we have analyzed the far-UV \textrm{C}~\textsc{iii}]$\lambda\lambda$1907, 1909 emission doublet in the LRIS spectra of a sample of 184 $z\sim$1 star-forming galaxies with \textrm{C}~\textsc{iii}] coverage, and list our key results below:

1. We have found that only $\sim20\%$ of the galaxies in the full sample (40 out of 184) have $>3\sigma$ \textrm{C}~\textsc{iii}] detections, with a median EW of $1.3\mbox{\AA}$, which is significantly weaker than the typical values observed so far at $z>6$. 

2. We have further investigated the correlations between the \textrm{C}~\textsc{iii}] strength and galaxy properties. By studying both individual \textrm{C}~\textsc{iii}] detections and composite spectra binned according to $B$-band absolute magnitude, $\ub$ color, stellar mass, SFR and sSFR, we have discovered that larger \textrm{C}~\textsc{iii}] EW appears in fainter, bluer and lower-mass systems, in which the gas metallicity and dust extinction are likely to be lower as well.

3. We have also explored the connections between \textrm{C}~\textsc{iii}] and near-UV emission transitions, specifically \textrm{Fe}~\textsc{ii}* and \textrm{Mg}~\textsc{ii}, using composite spectra created according to galaxy properties. While we observed no significant dependence of \textrm{C}~\textsc{iii}] on \textrm{Fe}~\textsc{ii}* in terms of strength, \textrm{C}~\textsc{iii}] EW increases with increasing inferred \textrm{Mg}~\textsc{ii} emission. The behavior of \textrm{C}~\textsc{iii}], \textrm{Fe}~\textsc{ii}*, and \textrm{Mg}~\textsc{ii} can be explained by different source functions for these features: there are net sources of nebular \textrm{C}~\textsc{iii}] and \textrm{Mg}~\textsc{ii} photons in \textrm{H}~\textsc{ii} regions, whereas \textrm{Fe}~\textsc{ii}* emission originates in circumgalactic, perhaps outflowing gas at larger radii, representing the redistribution of the stellar continuum (i.e., no underlying emission feature). 

4. Finally, we examined the effect metallicity has on the strength of \textrm{C}~\textsc{iii}] by adopting [\textrm{O}~\textsc{iii}]/$\mbox{H}\beta$ as a crude metallicity diagnostic. From the composite spectra created out of 26 $z<0.9$ objects with [\textrm{O}~\textsc{iii}] and $\mbox{H}\beta$ coverage in the DEEP2 DEIMOS spectra, we indirectly inferred stronger \textrm{C}~\textsc{iii}] emission at lower metallicity, based on the variation of \textrm{Mg}~\textsc{ii} absorption with [\textrm{O}~\textsc{iii}]/$\mbox{H}\beta$. This observational result is consistent with models showing that the harder ionizing spectrum and higher nebular gas temperature at lower oxygen abundance lead to enhanced \textrm{C}~\textsc{iii}] emission by producing more ionizing photons and suppressing metal cooling \citep{Jaskot2016}. 

5. We concluded that higher metallicity and older stellar population ages are probably the major causes of the lower typical \textrm{C}~\textsc{iii}] EW in our $z\sim1$ sample, compared with \textrm{C}~\textsc{iii}] detections at $z>6$.

Extreme \textrm{C}~\textsc{iii}] emitters at $z>6$ (with an average \textrm{C}~\textsc{iii}] EW $>10\mbox{\AA}$) are important for our understanding of the physical conditions in galaxies at early times. However, due to the low $S/N$ and the generally poor detection of metal lines, searching for such galaxies is exceptionally challenging. An alternative way to learn about these high-redshift \textrm{C}~\textsc{iii}] emitters is by using low-redshift analogs. With low-redshift observations, we can potentially obtain detailed information, with higher $S/N$, on the physical properties (e.g., metallicity, ionization parameter, dust extinction) from rest-UV and optical spectroscopy, and infer the stellar populations from SED fits using multi-wavelength photometry.

Despite the lensed sample from \citet{Stark2014}, the abundance of galaxies with strong rest-UV emission is especially low at $z\sim1-3$. Therefore, our main aim for future observations is to find the ``missing" \textrm{C}~\textsc{iii}] emitters within this redshift range. We will prioritize objects for rest-UV spectroscopy that show blue color (see discussion in Section \ref{sec:galprop}) and large [\textrm{O}~\textsc{iii}] EW, as the strength of \textrm{C}~\textsc{iii}] has found to be positively correlated with that of [\textrm{O}~\textsc{iii}] \citep{Stark2014,Berg2016,Jaskot2016}. In addition, full coverage of rest-optical features (e.g., [\textrm{O}~\textsc{ii}], [\textrm{O}~\textsc{iii}] and $\mbox{H}\beta$) is essential for confirming the correlation of \textrm{C}~\textsc{iii}] EW and metallicity at this redshift. By studying the analogs of high-redshift extreme \textrm{C}~\textsc{iii}] emitters, we will gain a better understanding of the physical conditions that give rise to this nebular feature and the nature of the sources that contribute to reionization within the first billion years of cosmic time.

\acknowledgements We thank Benjamin Weiner, Charles Steidel, Danielle Berg, and Jane Rigby for providing additional data and line measurements, and Anne Jaskot for sharing modeling results in advance of publication. We acknowledge support from the David $\&$ Lucile Packard Foundation (A.E.S. and C.L.M.),
and the National Science Foundation through grants AST-0808161 and AST-1109288 (C.L.M.) and CAREER award AST-1055081 (A.L.C.).
We are grateful to the DEEP2 and AEGIS teams for providing both the galaxy sample and ancillary data on galaxy properties.
We wish to extend special thanks to those of Hawaiian ancestry on
whose sacred mountain we are privileged to be guests. Without their generous hospitality, most
of the observations presented herein would not have been possible.

\bibliographystyle{apj}
\nocite{*}
\bibliography{ms}

\begin{thebibliography}{}
\expandafter\ifx\csname natexlab\endcsname\relax\def\natexlab#1{#1}\fi

\bibitem[{{Akerman} {et~al.}(2004){Akerman}, {Carigi}, {Nissen}, {Pettini}, \&
  {Asplund}}]{Akerman2004}
{Akerman}, C.~J., {Carigi}, L., {Nissen}, P.~E., {Pettini}, M., \& {Asplund},
  M. 2004, \aap, 414, 931

\bibitem[{{Baum}(1962)}]{Baum1962}
{Baum}, W.~A. 1962, in IAU Symposium, Vol.~15, Problems of Extra-Galactic
  Research, ed. G.~C. {McVittie}, 390

\bibitem[{{Bayliss} {et~al.}(2014){Bayliss}, {Rigby}, {Sharon}, {Wuyts},
  {Florian}, {Gladders}, {Johnson}, \& {Oguri}}]{Bayliss2014}
{Bayliss}, M.~B., {Rigby}, J.~R., {Sharon}, K., {et~al.} 2014, \apj, 790, 144

\bibitem[{{Berg} {et~al.}(2016){Berg}, {Skillman}, {Henry}, {Erb}, \&
  {Carigi}}]{Berg2016}
{Berg}, D.~A., {Skillman}, E.~D., {Henry}, R.~B.~C., {Erb}, D.~K., \& {Carigi},
  L. 2016, \apj, 827, 126

\bibitem[{{Bruzual} \& {Charlot}(2003)}]{Bruzual2003}
{Bruzual}, G., \& {Charlot}, S. 2003, \mnras, 344, 1000

\bibitem[{{Bundy} {et~al.}(2006){Bundy}, {Ellis}, {Conselice}, {Taylor},
  {Cooper}, {Willmer}, {Weiner}, {Coil}, {Noeske}, \& {Eisenhardt}}]{Bundy2006}
{Bundy}, K., {Ellis}, R.~S., {Conselice}, C.~J., {et~al.} 2006, \apj, 651, 120

\bibitem[{{Caruana} {et~al.}(2014){Caruana}, {Bunker}, {Wilkins}, {Stanway},
  {Lorenzoni}, {Jarvis}, \& {Ebert}}]{Caruana2014}
{Caruana}, J., {Bunker}, A.~J., {Wilkins}, S.~M., {et~al.} 2014, \mnras, 443,
  2831

\bibitem[{{Chabrier}(2003)}]{Chabrier2003}
{Chabrier}, G. 2003, \pasp, 115, 763

\bibitem[{{de Barros} {et~al.}(2016){de Barros}, {Vanzella}, {Amor{\'{\i}}n},
  {Castellano}, {Siana}, {Grazian}, {Suh}, {Balestra}, {Vignali}, {Verhamme},
  {Zamorani}, {Mignoli}, {Hasinger}, {Comastri}, {Pentericci},
  {P{\'e}rez-Montero}, {Fontana}, {Giavalisco}, \& {Gilli}}]{Debarros2016}
{de Barros}, S., {Vanzella}, E., {Amor{\'{\i}}n}, R., {et~al.} 2016, \aap, 585,
  A51

\bibitem[{{Du} {et~al.}(2016){Du}, {Shapley}, {Martin}, \& {Coil}}]{Du2016}
{Du}, X., {Shapley}, A.~E., {Martin}, C.~L., \& {Coil}, A.~L. 2016, \apj, 829,
  64

\bibitem[{{Erb} {et~al.}(2010){Erb}, {Pettini}, {Shapley}, {Steidel}, {Law}, \&
  {Reddy}}]{Erb2010}
{Erb}, D.~K., {Pettini}, M., {Shapley}, A.~E., {et~al.} 2010, \apj, 719, 1168

\bibitem[{{Erb} {et~al.}(2012){Erb}, {Quider}, {Henry}, \& {Martin}}]{Erb2012}
{Erb}, D.~K., {Quider}, A.~M., {Henry}, A.~L., \& {Martin}, C.~L. 2012, \apj,
  759, 26

\bibitem[{{Erb} {et~al.}(2006){Erb}, {Shapley}, {Pettini}, {Steidel}, {Reddy},
  \& {Adelberger}}]{Erb2006}
{Erb}, D.~K., {Shapley}, A.~E., {Pettini}, M., {et~al.} 2006, \apj, 644, 813

\bibitem[{{Faber} {et~al.}(2003){Faber}, {Phillips}, {Kibrick}, {Alcott},
  {Allen}, {Burrous}, {Cantrall}, {Clarke}, {Coil}, {Cowley}, {Davis}, {Deich},
  {Dietsch}, {Gilmore}, {Harper}, {Hilyard}, {Lewis}, {McVeigh}, {Newman},
  {Osborne}, {Schiavon}, {Stover}, {Tucker}, {Wallace}, {Wei}, {Wirth}, \&
  {Wright}}]{Faber2003}
{Faber}, S.~M., {Phillips}, A.~C., {Kibrick}, R.~I., {et~al.} 2003, in
  \procspie, Vol. 4841, Instrument Design and Performance for Optical/Infrared
  Ground-based Telescopes, ed. M.~{Iye} \& A.~F.~M. {Moorwood}, 1657--1669

\bibitem[{{Faber} {et~al.}(2007){Faber}, {Willmer}, {Wolf}, {Koo}, {Weiner},
  {Newman}, {Im}, {Coil}, {Conroy}, {Cooper}, {Davis}, {Finkbeiner}, {Gerke},
  {Gebhardt}, {Groth}, {Guhathakurta}, {Harker}, {Kaiser}, {Kassin},
  {Kleinheinrich}, {Konidaris}, {Kron}, {Lin}, {Luppino}, {Madgwick},
  {Meisenheimer}, {Noeske}, {Phillips}, {Sarajedini}, {Schiavon}, {Simard},
  {Szalay}, {Vogt}, \& {Yan}}]{Faber2007}
{Faber}, S.~M., {Willmer}, C.~N.~A., {Wolf}, C., {et~al.} 2007, \apj, 665, 265

\bibitem[{{Ferland} {et~al.}(1998){Ferland}, {Korista}, {Verner}, {Ferguson},
  {Kingdon}, \& {Verner}}]{Ferland1998}
{Ferland}, G.~J., {Korista}, K.~T., {Verner}, D.~A., {et~al.} 1998, \pasp, 110,
  761

\bibitem[{{Garnett} {et~al.}(1995){Garnett}, {Skillman}, {Dufour}, {Peimbert},
  {Torres-Peimbert}, {Terlevich}, {Terlevich}, \& {Shields}}]{Garnett1995}
{Garnett}, D.~R., {Skillman}, E.~D., {Dufour}, R.~J., {et~al.} 1995, \apj, 443,
  64

\bibitem[{{Giavalisco} {et~al.}(1996){Giavalisco}, {Koratkar}, \&
  {Calzetti}}]{Gia1996}
{Giavalisco}, M., {Koratkar}, A., \& {Calzetti}, D. 1996, \apj, 466, 831

\bibitem[{{Gutkin} {et~al.}(2016){Gutkin}, {Charlot}, \&
  {Bruzual}}]{Gutkin2016}
{Gutkin}, J., {Charlot}, S., \& {Bruzual}, G. 2016, \mnras, 462, 1757

\bibitem[{{Heckman} {et~al.}(1998){Heckman}, {Robert}, {Leitherer}, {Garnett},
  \& {van der Rydt}}]{Heckman1998}
{Heckman}, T.~M., {Robert}, C., {Leitherer}, C., {Garnett}, D.~R., \& {van der
  Rydt}, F. 1998, \apj, 503, 646

\bibitem[{{Henry} {et~al.}(2000){Henry}, {Edmunds}, \&
  {K{\"o}ppen}}]{Henry2000}
{Henry}, R.~B.~C., {Edmunds}, M.~G., \& {K{\"o}ppen}, J. 2000, \apj, 541, 660

\bibitem[{{Jaskot} \& {Ravindranath}(2016)}]{Jaskot2016}
{Jaskot}, A.~E., \& {Ravindranath}, S. 2016, \apj, 833, 136

\bibitem[{{Jones} {et~al.}(2015){Jones}, {Martin}, \& {Cooper}}]{Jones2015}
{Jones}, T., {Martin}, C., \& {Cooper}, M.~C. 2015, \apj, 813, 126

\bibitem[{{Kornei} {et~al.}(2012){Kornei}, {Shapley}, {Martin}, {Coil}, {Lotz},
  {Schiminovich}, {Bundy}, \& {Noeske}}]{Kornei2012}
{Kornei}, K.~A., {Shapley}, A.~E., {Martin}, C.~L., {et~al.} 2012, \apj, 758,
  135

\bibitem[{{Kornei} {et~al.}(2013){Kornei}, {Shapley}, {Martin}, {Coil}, {Lotz},
  \& {Weiner}}]{Kornei2013}
---. 2013, \apj, 774, 50

\bibitem[{{Leitherer} {et~al.}(2011){Leitherer}, {Tremonti}, {Heckman}, \&
  {Calzetti}}]{Leitherer2011}
{Leitherer}, C., {Tremonti}, C.~A., {Heckman}, T.~M., \& {Calzetti}, D. 2011,
  \aj, 141, 37

\bibitem[{{Madau} \& {Dickinson}(2014)}]{Madau2014}
{Madau}, P., \& {Dickinson}, M. 2014, \araa, 52, 415

\bibitem[{{Maiolino} {et~al.}(2008){Maiolino}, {Nagao}, {Grazian}, {Cocchia},
  {Marconi}, {Mannucci}, {Cimatti}, {Pipino}, {Ballero}, {Calura}, {Chiappini},
  {Fontana}, {Granato}, {Matteucci}, {Pastorini}, {Pentericci}, {Risaliti},
  {Salvati}, \& {Silva}}]{Maiolino2008}
{Maiolino}, R., {Nagao}, T., {Grazian}, A., {et~al.} 2008, \aap, 488, 463

\bibitem[{{Mannucci} {et~al.}(2010){Mannucci}, {Cresci}, {Maiolino}, {Marconi},
  \& {Gnerucci}}]{Mannucci2010}
{Mannucci}, F., {Cresci}, G., {Maiolino}, R., {Marconi}, A., \& {Gnerucci}, A.
  2010, \mnras, 408, 2115

\bibitem[{{Markwardt}(2009)}]{Mark2009}
{Markwardt}, C.~B. 2009, in Astronomical Society of the Pacific Conference
  Series, Vol. 411, Astronomical Data Analysis Software and Systems XVIII, ed.
  D.~A. {Bohlender}, D.~{Durand}, \& P.~{Dowler}, 251

\bibitem[{{Martin} {et~al.}(2012){Martin}, {Shapley}, {Coil}, {Kornei},
  {Bundy}, {Weiner}, {Noeske}, \& {Schiminovich}}]{Martin2012}
{Martin}, C.~L., {Shapley}, A.~E., {Coil}, A.~L., {et~al.} 2012, \apj, 760, 127

\bibitem[{{Martin} {et~al.}(2013){Martin}, {Shapley}, {Coil}, {Kornei},
  {Murray}, \& {Pancoast}}]{Martin2013}
---. 2013, \apj, 770, 41

\bibitem[{{Newman} {et~al.}(2013){Newman}, {Cooper}, {Davis}, {Faber}, {Coil},
  {Guhathakurta}, {Koo}, {Phillips}, {Conroy}, {Dutton}, {Finkbeiner}, {Gerke},
  {Rosario}, {Weiner}, {Willmer}, {Yan}, {Harker}, {Kassin}, {Konidaris},
  {Lai}, {Madgwick}, {Noeske}, {Wirth}, {Connolly}, {Kaiser}, {Kirby},
  {Lemaux}, {Lin}, {Lotz}, {Luppino}, {Marinoni}, {Matthews}, {Metevier}, \&
  {Schiavon}}]{Newman2013}
{Newman}, J.~A., {Cooper}, M.~C., {Davis}, M., {et~al.} 2013, \apjs, 208, 5

\bibitem[{{Noeske} {et~al.}(2007){Noeske}, {Weiner}, {Faber}, {Papovich},
  {Koo}, {Somerville}, {Bundy}, {Conselice}, {Newman}, {Schiminovich}, {Le
  Floc'h}, {Coil}, {Rieke}, {Lotz}, {Primack}, {Barmby}, {Cooper}, {Davis},
  {Ellis}, {Fazio}, {Guhathakurta}, {Huang}, {Kassin}, {Martin}, {Phillips},
  {Rich}, {Small}, {Willmer}, \& {Wilson}}]{Noeske2007}
{Noeske}, K.~G., {Weiner}, B.~J., {Faber}, S.~M., {et~al.} 2007, \apjl, 660,
  L43

\bibitem[{{Oke} {et~al.}(1995){Oke}, {Cohen}, {Carr}, {Cromer}, {Dingizian},
  {Harris}, {Labrecque}, {Lucinio}, {Schaal}, {Epps}, \& {Miller}}]{Oke1995}
{Oke}, J.~B., {Cohen}, J.~G., {Carr}, M., {et~al.} 1995, \pasp, 107, 375

\bibitem[{{Osterbrock} \& {Ferland}(2006)}]{Oster2006}
{Osterbrock}, D.~E., \& {Ferland}, G.~J. 2006, {Astrophysics of gaseous nebulae
  and active galactic nuclei}

\bibitem[{{Pentericci} {et~al.}(2014){Pentericci}, {Vanzella}, {Fontana},
  {Castellano}, {Treu}, {Mesinger}, {Dijkstra}, {Grazian}, {Brada{\v c}},
  {Conselice}, {Cristiani}, {Dunlop}, {Galametz}, {Giavalisco}, {Giallongo},
  {Koekemoer}, {McLure}, {Maiolino}, {Paris}, \& {Santini}}]{Pentericci2014}
{Pentericci}, L., {Vanzella}, E., {Fontana}, A., {et~al.} 2014, \apj, 793, 113

\bibitem[{{Rigby} {et~al.}(2015){Rigby}, {Bayliss}, {Gladders}, {Sharon},
  {Wuyts}, {Dahle}, {Johnson}, \& {Pe{\~n}a-Guerrero}}]{Rigby2015}
{Rigby}, J.~R., {Bayliss}, M.~B., {Gladders}, M.~D., {et~al.} 2015, \apjl, 814,
  L6

\bibitem[{{Rix} {et~al.}(2004){Rix}, {Pettini}, {Leitherer}, {Bresolin},
  {Kudritzki}, \& {Steidel}}]{Rix2004}
{Rix}, S.~A., {Pettini}, M., {Leitherer}, C., {et~al.} 2004, \apj, 615, 98

\bibitem[{{Rubin} {et~al.}(2011){Rubin}, {Prochaska}, {M{\'e}nard}, {Murray},
  {Kasen}, {Koo}, \& {Phillips}}]{Rubin2011}
{Rubin}, K.~H.~R., {Prochaska}, J.~X., {M{\'e}nard}, B., {et~al.} 2011, \apj,
  728, 55

\bibitem[{{Salim} {et~al.}(2007){Salim}, {Rich}, {Charlot}, {Brinchmann},
  {Johnson}, {Schiminovich}, {Seibert}, {Mallery}, {Heckman}, {Forster},
  {Friedman}, {Martin}, {Morrissey}, {Neff}, {Small}, {Wyder}, {Bianchi},
  {Donas}, {Lee}, {Madore}, {Milliard}, {Szalay}, {Welsh}, \& {Yi}}]{Salim2007}
{Salim}, S., {Rich}, R.~M., {Charlot}, S., {et~al.} 2007, \apjs, 173, 267

\bibitem[{{Schenker} {et~al.}(2014){Schenker}, {Ellis}, {Konidaris}, \&
  {Stark}}]{Schenker2014}
{Schenker}, M.~A., {Ellis}, R.~S., {Konidaris}, N.~P., \& {Stark}, D.~P. 2014,
  \apj, 795, 20

\bibitem[{{Seibert} {et~al.}(2005){Seibert}, {Martin}, {Heckman}, {Buat},
  {Hoopes}, {Barlow}, {Bianchi}, {Byun}, {Donas}, {Forster}, {Friedman},
  {Jelinsky}, {Lee}, {Madore}, {Malina}, {Milliard}, {Morrissey}, {Neff},
  {Rich}, {Schiminovich}, {Siegmund}, {Small}, {Szalay}, {Welsh}, \&
  {Wyder}}]{Seibert2005}
{Seibert}, M., {Martin}, D.~C., {Heckman}, T.~M., {et~al.} 2005, \apjl, 619,
  L55

\bibitem[{{Shapley} {et~al.}(2003){Shapley}, {Steidel}, {Pettini}, \&
  {Adelberger}}]{Shapley2003}
{Shapley}, A.~E., {Steidel}, C.~C., {Pettini}, M., \& {Adelberger}, K.~L. 2003,
  \apj, 588, 65

\bibitem[{{Stanway} {et~al.}(2016){Stanway}, {Eldridge}, \&
  {Becker}}]{Stanway2016}
{Stanway}, E.~R., {Eldridge}, J.~J., \& {Becker}, G.~D. 2016, \mnras, 456, 485

\bibitem[{{Stark} {et~al.}(2014){Stark}, {Richard}, {Siana}, {Charlot},
  {Freeman}, {Gutkin}, {Wofford}, {Robertson}, {Amanullah}, {Watson}, \&
  {Milvang-Jensen}}]{Stark2014}
{Stark}, D.~P., {Richard}, J., {Siana}, B., {et~al.} 2014, \mnras, 445, 3200

\bibitem[{{Stark} {et~al.}(2015){Stark}, {Richard}, {Charlot}, {Cl{\'e}ment},
  {Ellis}, {Siana}, {Robertson}, {Schenker}, {Gutkin}, \&
  {Wofford}}]{Stark2015}
{Stark}, D.~P., {Richard}, J., {Charlot}, S., {et~al.} 2015, \mnras, 450, 1846

\bibitem[{{Stark} {et~al.}(2017){Stark}, {Ellis}, {Charlot}, {Chevallard},
  {Tang}, {Belli}, {Zitrin}, {Mainali}, {Gutkin}, {Vidal-Garc{\'{\i}}a},
  {Bouwens}, \& {Oesch}}]{Stark2016}
{Stark}, D.~P., {Ellis}, R.~S., {Charlot}, S., {et~al.} 2017, \mnras, 464, 469

\bibitem[{{Steidel} {et~al.}(2010){Steidel}, {Erb}, {Shapley}, {Pettini},
  {Reddy}, {Bogosavljevi{\'c}}, {Rudie}, \& {Rakic}}]{Steidel2010}
{Steidel}, C.~C., {Erb}, D.~K., {Shapley}, A.~E., {et~al.} 2010, \apj, 717, 289

\bibitem[{{Steidel} {et~al.}(1996){Steidel}, {Giavalisco}, {Dickinson}, \&
  {Adelberger}}]{Steidel1996}
{Steidel}, C.~C., {Giavalisco}, M., {Dickinson}, M., \& {Adelberger}, K.~L.
  1996, \aj, 112, 352

\bibitem[{{Steidel} {et~al.}(2004){Steidel}, {Shapley}, {Pettini},
  {Adelberger}, {Erb}, {Reddy}, \& {Hunt}}]{Steidel2004}
{Steidel}, C.~C., {Shapley}, A.~E., {Pettini}, M., {et~al.} 2004, \apj, 604,
  534

\bibitem[{{Steidel} {et~al.}(2016){Steidel}, {Strom}, {Pettini}, {Rudie},
  {Reddy}, \& {Trainor}}]{Steidel2016}
{Steidel}, C.~C., {Strom}, A.~L., {Pettini}, M., {et~al.} 2016, \apj, 826, 159

\bibitem[{{Tilvi} {et~al.}(2014){Tilvi}, {Papovich}, {Finkelstein}, {Long},
  {Song}, {Dickinson}, {Ferguson}, {Koekemoer}, {Giavalisco}, \&
  {Mobasher}}]{Tilvi2014}
{Tilvi}, V., {Papovich}, C., {Finkelstein}, S.~L., {et~al.} 2014, \apj, 794, 5

\bibitem[{{Tremonti} {et~al.}(2004){Tremonti}, {Heckman}, {Kauffmann},
  {Brinchmann}, {Charlot}, {White}, {Seibert}, {Peng}, {Schlegel}, {Uomoto},
  {Fukugita}, \& {Brinkmann}}]{Tremonti2004}
{Tremonti}, C.~A., {Heckman}, T.~M., {Kauffmann}, G., {et~al.} 2004, \apj, 613,
  898

\bibitem[{{Treu} {et~al.}(2013){Treu}, {Schmidt}, {Trenti}, {Bradley}, \&
  {Stiavelli}}]{Treu2013}
{Treu}, T., {Schmidt}, K.~B., {Trenti}, M., {Bradley}, L.~D., \& {Stiavelli},
  M. 2013, \apjl, 775, L29

\bibitem[{{Treu} {et~al.}(2012){Treu}, {Trenti}, {Stiavelli}, {Auger}, \&
  {Bradley}}]{Treu2012}
{Treu}, T., {Trenti}, M., {Stiavelli}, M., {Auger}, M.~W., \& {Bradley}, L.~D.
  2012, \apj, 747, 27

\bibitem[{{Vanzella} {et~al.}(2011){Vanzella}, {Pentericci}, {Fontana},
  {Grazian}, {Castellano}, {Boutsia}, {Cristiani}, {Dickinson}, {Gallozzi},
  {Giallongo}, {Giavalisco}, {Maiolino}, {Moorwood}, {Paris}, \&
  {Santini}}]{Vanzella2011}
{Vanzella}, E., {Pentericci}, L., {Fontana}, A., {et~al.} 2011, \apjl, 730, L35

\bibitem[{{Vanzella} {et~al.}(2016){Vanzella}, {De Barros}, {Cupani}, {Karman},
  {Gronke}, {Balestra}, {Coe}, {Mignoli}, {Brusa}, {Calura}, {Caminha},
  {Caputi}, {Castellano}, {Christensen}, {Comastri}, {Cristiani}, {Dijkstra},
  {Fontana}, {Giallongo}, {Giavalisco}, {Gilli}, {Grazian}, {Grillo},
  {Koekemoer}, {Meneghetti}, {Nonino}, {Pentericci}, {Rosati}, {Schaerer},
  {Verhamme}, {Vignali}, \& {Zamorani}}]{Vanzella2016}
{Vanzella}, E., {De Barros}, S., {Cupani}, G., {et~al.} 2016, \apjl, 821, L27

\bibitem[{{Weiner} {et~al.}(2009){Weiner}, {Coil}, {Prochaska}, {Newman},
  {Cooper}, {Bundy}, {Conselice}, {Dutton}, {Faber}, {Koo}, {Lotz}, {Rieke}, \&
  {Rubin}}]{Weiner2009}
{Weiner}, B.~J., {Coil}, A.~L., {Prochaska}, J.~X., {et~al.} 2009, \apj, 692,
  187

\bibitem[{{Willmer} {et~al.}(2006){Willmer}, {Faber}, {Koo}, {Weiner},
  {Newman}, {Coil}, {Connolly}, {Conroy}, {Cooper}, {Davis}, {Finkbeiner},
  {Gerke}, {Guhathakurta}, {Harker}, {Kaiser}, {Kassin}, {Konidaris}, {Lin},
  {Luppino}, {Madgwick}, {Noeske}, {Phillips}, \& {Yan}}]{Willmer2006}
{Willmer}, C.~N.~A., {Faber}, S.~M., {Koo}, D.~C., {et~al.} 2006, \apj, 647,
  853

\bibitem[{{Zahid} {et~al.}(2013){Zahid}, {Geller}, {Kewley}, {Hwang},
  {Fabricant}, \& {Kurtz}}]{Zahid2013}
{Zahid}, H.~J., {Geller}, M.~J., {Kewley}, L.~J., {et~al.} 2013, \apjl, 771,
  L19

\bibitem[{{Zakamska} {et~al.}(2003){Zakamska}, {Strauss}, {Krolik}, {Collinge},
  {Hall}, {Hao}, {Heckman}, {Ivezi{\'c}}, {Richards}, {Schlegel}, {Schneider},
  {Strateva}, {Vanden Berk}, {Anderson}, \& {Brinkmann}}]{Zakamska2003}
{Zakamska}, N.~L., {Strauss}, M.~A., {Krolik}, J.~H., {et~al.} 2003, \aj, 126,
  2125

\bibitem[{{Zitrin} {et~al.}(2015){Zitrin}, {Ellis}, {Belli}, \&
  {Stark}}]{Zitrin2015}
{Zitrin}, A., {Ellis}, R.~S., {Belli}, S., \& {Stark}, D.~P. 2015, \apjl, 805,
  L7

\end{thebibliography}
\end{CJK}
\end{document}